\documentclass[preprint,showpacs,preprintnumbers,amsmath,amssymb]{revtex4}
\usepackage{graphicx,epsfig,dcolumn,bm,epic,eepic,float}%
\usepackage{amsmath}
\usepackage{latexsym}
\usepackage{color}
\usepackage{cancel}
\usepackage{makeidx,shortvrb,latexsym}
\usepackage{slashed}
\begin{document}
\unitlength 1 cm
\newcommand{\nn}{\nonumber}
\newcommand{\vk}{\vec k}
\newcommand{\vp}{\vec p}
\newcommand{\vq}{\vec q}
\newcommand{\vkp}{\vec {k'}}
\newcommand{\vpp}{\vec {p'}}
\newcommand{\vqp}{\vec {q'}}
\newcommand{\bk}{{\bf k}}
\newcommand{\bp}{{\bf p}}
\newcommand{\bq}{{\bf q}}
\newcommand{\br}{{\bf r}}
\newcommand{\bR}{{\bf R}}
\newcommand{\up}{\uparrow}
\newcommand{\down}{\downarrow}
\newcommand{\fns}{\footnotesize}
\newcommand{\ns}{\normalsize}
\newcommand{\cdag}{c^{\dagger}}
\title {A detail study of the LHC and TEVATRON hadron-hadron prompt-photon pair production
experiments in the angular ordering constraint  $k_t$-factorization
approaches}
\author{$M. \; Modarres$ }
\altaffiliation {Corresponding author, Email:
mmodares@ut.ac.ir,Tel:+98-21-61118645, Fax:+98-21-88004781.}
\author{$R. \; Aminzadeh\;Nik$ }
\author{$R.\; Kord\; Valeshbadi$}
\affiliation {Department of Physics, University of $Tehran$,
1439955961, $Tehran$, Iran.}
\author{H. Hosseinkhani }\affiliation{Plasma and Fusion Research School, Nuclear
Science and Technology Research Institute, 14395-836 Tehran, Iran.}
\author{$N. Olanj$}\affiliation {Physics Department,
Faculty of Science, $Bu$-$Ali Sina$ University, 65178, $Hamedan$,
Iran.}
\begin{abstract}
In the present work, which is based on the $k_t$-factorization
framework, it is intended to make a detail study of the isolated
prompt-photon pairs (IPPP) production in the high-energy inelastic
hadron-hadron collisions differential cross section.  The two
scheme-dependent unintegrated parton distribution functions (UPDF)
in which the angular ordering constraints (AOC) are imposed, namely
the Kimber-Martin-Ryskin (KMR) and the Martin-Ryskin-Watt (MRW)
approaches, in the leading and the next-to-leading orders (LO and
NLO) are considered, respectively. These two prescriptions (KMR and
MRW) utilize the phenomenological parton distribution functions
(PDF) libraries of Martin et al, i.e. the
 MMHT2014. The computations are performed
in accordance with the initial dynamics of  latest existing
experimental reports of the D0, CDF, CMS and ATLAS collaborations
and the different experimental constraints. It is
  shown that above  frameworks are capable of producing acceptable
results, compared to the experimental data, the pQCD and some Monte
Carlo calculations (i.e. $2\gamma$NNLO, SHERPA, DIPHOX and RESBOS).
It is also concluded that the KMR framework produces better results
in the higher center-of-mass energies, while the same thing can be
argued about the LO-MRW prescription in lower energies.
Additionally, these two schemes show different behavior in the
regions where the fragmentation and higher pQCD effects become
important. A clear prediction for the various shoulders and tails
which were detected experimentally are observed and discussed  in
the present theoretical approaches. The possible double countings
between 2$\rightarrow$2 and 2$\rightarrow$3 processes are studied.
Finally, in agreement to the work of Golec-Biernat and Stasto, it is
shown that there is not any dispute about the application of the AOC
and the cut off, in the above prescriptions at least in the
calculation of the various IPPP differential cross sections.
\end{abstract}
 \maketitle
\section{Introduction}
The study of photon pair production plays an important role in the
investigation of: (1) the perturbative quantum chromodynamics (pQCD)
and (2) better observation of the Higgs boson's decay to diphotons,
as well as (3) some theories, which extended beyond the standard
model and should give some predictions regarding  the new phenomena
in the fundamental particle physics \cite{collins}. Many
experimental efforts at the LHC and  TEVATRON colliders  have been
performed to explore the  physics of these regions, e.g. the D0,
CDF, CMS and ATLAS collaborations
\cite{D010,D013,CDF11,CDF13,CMS11,CMS14,ATLAS10,ATLAS13}. These
investigations are probing different channels and exploring
different aspects of the above  subjects, such as producing
differential cross-section of the photon pair production as a
function of the azimuthal separation angle between the photon pair
in the laboratory frame ($\Delta\phi_{\gamma \gamma} $) and the
transverse momenta of the  photon pair ($p_{t,\gamma\gamma} $). They
are particularly useful to study the higher order pQCD and the
fragmentation effects \cite{ATLAS13}. Other observable, namely the
photon pair invariant-mass ($M_{\gamma \gamma}$) and the polar angle
of the highest  photon-transfer-energy-momentum in the Collins-Soper
isolated prompt photon pair (IPPP) rest frame ($cos\theta_{\gamma
\gamma}^{*}$), are also powerful tools to investigate the spin of
the  photon pair resonances \cite{ATLAS13}. The experimental
collaborations conventionally use some parton level Monte-Carlo
programs, e.g. RESBOS, DIPHOX,  $2\gamma$NNLO and SHERPA
\cite{RESBOS,DIPHOX,NNLO,sherpa}, to test the pQCD theory against
their data. Also, in the recent years, (2016), a new article
  based on MCFM  program, (Monte Carlo for FeMtobarn), which uses the collinear
  factorization formalism,
with NNLO accuracy was published that its result has good agreement
with the available data \cite{new}.

The RESBOS Monte-Carlo event generator, provides the next-to-leading
order (NLO) level pQCD predictions for the IPPP with the soft gluon
re-summation which can   include the
 single photon fragmentation \cite{tomas} as well. The DIPHOX
Monte-Carlo event generator performs   the IPPP production at the
NLO pQCD level, in which  the single and double fragmentation
contributions \cite{tomas} are also included. The $2\gamma$NNLO is
developed to include the full next-to-next-to-leading order (NNLO)
pQCD, without considering fragmentation contributions \cite{NNLO}.
Another choice is  the SHERPA \cite{sherpa}  Monte Carlo event
generator, that could  simulate  the high-energy reactions of
particles in the hadron-hadron collisions.

To perform such a analysis one usually needs parton distribution
functions (PDF), $a(x,Q^2)$, or unintegrated PDF (UPDF),
$f(x,k_t,Q^2$), see the references
\cite{DGLAP1,DGLAP2,DGLAP3,DGLAP4,BFKL1,BFKL2,BFKL3,BFKL4,BFKL5,CCFM1,CCFM2,CCFM3,CCFM4,CCFM5,Q-CCFMI,Q-CCFMII}
and the section III. Note that $x$, $Q^2$ and $k_t$ are the Bjorken
scale,the hard scale and the parton transverse momentum.

In the most of the recent theories, which explore the domain beyond
the standard model, the photon is expected to be present at the
final states. Therefore, we expect to see some features of these
theories in the existing experimental data. Consequently, a detailed
analysis over the experimental data is vital, to determine the exact
contributions out of the standard model, in order to justify or
reject such theories \cite{Lip0}.

Regarding the complication and the weakness of  different
prescriptions
\cite{DGLAP1,DGLAP2,DGLAP3,DGLAP4,BFKL1,BFKL2,BFKL3,BFKL4,BFKL5,CCFM1,CCFM2,CCFM3,CCFM4,CCFM5,Q-CCFMI,Q-CCFMII},
Martin et al \cite{KMR,MRW} defined the UPDF in the
$k_t$-factorization framework, in relation to the conventional PDF
\cite{MMHT}, through the identity,
\begin{equation}
    {xa(x,Q^2) \simeq \int^{Q^2} {{dk_t^2}\over{k_t^2}}} f(x,k^2_t,Q^2),
    \label{eq0065}
\end{equation}
and  developed the Kimber-Martin-Ryskin (KMR) and the
Martin-Ryskin-Watt (MRW) approaches \cite{KMR,MRW}. These formalisms
were analyzed thoroughly via the calculation of the proton structure
functions ($F_2$ and $F_L$) in the references
\cite{Mod1,Mod2,Mod3,Mod4,Mod5, Mod6,Mod7,FL}. Also, the
applications of    KMR and MRW frameworks   in the LO and the NLO
levels were investigated against the existing experimental data in
the references \cite{FL,FL-dipole,W/Z-NLO,W/Z-LHCb,Di-jet}, and some
successful results were achieved.

In the present work, we intend to study the production of the IPPP
in the high energy Hadron-Hadron collisions, in the frameworks of
KMR and MRW procedures. A primary investigation, using the KMR
$k_t$-factorization approach, was performed in the reference
\cite{LIPIPPP}, with some comparisons to the old data
\cite{D010,CDF11,CMS11,ATLAS10}, and some discrepancies especially
in the fragmentation regions,   were observed. To investigate this
problem, it is intended to use three different procedure via the
$k_t$-factorization formalism, by
 utilizing the UPDF of KMR, LO-MRW and NLO-MRW frameworks. Then the extracted results are compared with
the latest, as well as old, experimental data of the D0, CDF, CMS
and ATLAS collaborations in their respective dynamical
specifications
\cite{D010,D013,CDF11,CDF13,CMS11,CMS14,ATLAS10,ATLAS13} and other
theoretical approaches \cite{RESBOS,sherpa,NNLO,tomas,DIPHOX}
discussed above. It will be shown that the $k_t$-factorization
framework is reasonably capable of  describing of the high energy
experiments data for the IPPP production. We also discuss the
various advantages and disadvantages of the KMR and MRW
prescriptions in connection to each experiments conditions by
presenting a detail comparison. One of the main goals of our work is
to observe and analyze the effect of imposing different
visualizations of the AOC (embedded in different UPDF prescription
schemes) in the partonic dynamics that depends on the deriving
factors, (i.e. different experimental constraints), which are
assumed in the existing experimental data, such that to cover the
sensitive area to the fragmentation and the higher order pQCD
effects  (see also the sections  V and VI). On the other hand, very
recently there was some dispute about the application of the AOC and
the cut off in the KMR prescription \cite{re} which is different in
case of the MRW (note that for the KMR scheme the AOC is applied on
both quark and gluon radiations but this is not the case in the MRW
approaches (see the section III)). As it was discussed in the
reference \cite{re}, our calculations show
 that a qualitative agreements between the different schemes can be
achieve at least in the calculation of differential  cross sections
\cite{AMIN}. Beside these, the ambiguity about the fragmentation
region \cite{LIPIPPP} is considered by performing MRW-LO, which show
a better agreement with data at the fragmentation regions. On the
other hand the Guillet shoulder \cite{Guilet binoth} as well as new
shoulders are observed (see the section V). In the reference
\cite{LIPIPPP}, the
 valence quarks were only considered in the case     of $q^*$+$\bar{q}^*$ for $P\bar{P}$
collision and the see-quarks contributions (which is very small)
were ignored. Beside these, in  the PP collision the latter
contributions are sizable, in contrast to the reference
\cite{LIPIPPP} which again was not included. There are also some
other essential points which will be discussed in the end of section
II, IV and V.

In the above calculation, one should evaluate the off-shell
transition matrix elements.  Various formalisms were introduced to
calculate these off-shell matrix elements to insure the gauge
invariance and the satisfactions of the Ward identities
\cite{KUTAK1,KUTAK2,KUTAK3,KUTAK4,1p,saleev,saleev1,LIP2016}. The
off-shell matrix element violates the gauge invariant which is
necessary for the cross section calculation. In the references
\cite{1p,KUTAK4,2}, it was  shown that a suitable gauges for gluons
and photons polarizations lead to saving the gauge invariant of the
off-shell gluons matrix elements. The eikonal polarization is the
result of using axial gauge which is considered  in the current work
(see the section II). But the problem of gauge invariance violation
is still remained in all processes that the quarks are the incoming
off-shell legs. However,
 in
the small x limit and the large transverse momentum, using the
approximation made in the references  \cite{lip31,SPL}, the
off-shell matrix element   satisfies the gauge invariance
requirements. In this work, with the aforementioned constraints (the
small x limit and the large transverse momentum), we check
numerically the gauge invariant of each process individually.
However there are (i) reggeization methods to evaluate the off-shell
quark density matrix elements which are inherently satisfy the gauge
invariance in the all regions \cite{saleev,saleev1} or (ii)  the
method developed in the references \cite{KUTAK1} , in which by
modifying the vertexes and using auxiliary photons and quarks, an
off-shell quark matrix elements are produced, which satisfy the Ward
identity. To be insure about the above problems, in the section V we
check our result against those of reference \cite{saleev1}.

The possible double countings between 2$\rightarrow$2 and
2$\rightarrow$3 processes,  which were pointed out in the references
\cite{saleev1,LIP2011} as well as our previous work \cite{W/Z-LHCb},
will be discussed in the sections II, V and VI.

The others theoretical and the Monte Carlo calculation (as explained
above) are also presented against our results.  However, we
 should make this note that, as it was discussed in the reference
\cite{WattWZ}, the present approach should not be as good as pQCD
approaches, on the other hand, it is more simplistic.

In what follows,   the theoretical framework of the IPPP production
is presented in the section II. A brief introduction to the
$k_t$-factorization, and individually the KMR, LO-MRW and NLO-MRW
prescriptions, are presented in the section III. The section IV
contains a comprehensive description about the methods and the tools
for the calculation of the $k_t$-dependent cross-section of the IPPP
production   in the various proton-proton (or proton-antiproton)
inelastic collisions. The constraints of each experiment   are
discussed in the appendix A and finally, results, discussions and
 conclusions are given in the sections V and VI,
respectively.
\section{The Theoretical framework}
In the study of photon production, there exists two possible
categories; the prompt-photon and the non-prompt-photon. The first,
includes the fragmentation and the direct production of a photon
while the second is created in the processes of hadronic decay. In
this paper, we intend to base our calculations only on direct IPPP
production. In order to set the kinematics of the pair photon
production, we choose to work in the center-of-mass frame of the
initial  protons. So, we can set their four-momenta as: $$ P_1 =
{\sqrt{s} \over 2} (1,0,0,1), \;\;\; P_2 = {\sqrt{s} \over 2}
(1,0,0,-1)
    $$
where $s$ is the center-of-mass (CM) energy and  $P_1$ and $P_2$ are
the four-momenta of the colliding protons. One can write the total
cross-section for the production of the prompt-photons, summing over
all the contributing partonic sub-processes, i.e. $q^* + \bar{q}^*
\to \gamma + \gamma$, $q^*(\bar{q}^*) + g^* \to \gamma + \gamma +
q(\bar{q})$ and $g^* + g^* \to \gamma + \gamma$ \cite{LIPIPPP}.
Hence, it is required to write the four-momenta of the incoming
partons based on the four-momenta of the initial protons, using the
Sudakov decomposition assumption as,
    \begin{equation}
 {k}_i = x_i  {P}_i +  {k}_{i,t}, \;\;\;
    \label{eq22}
    \end{equation}
where  as we pointed out before, $k_{i,t}$ are the transverse
momenta of the $i^{th}$ partons and $x_i$ are the fraction of the
longitudinal momentum of the protons that are inherited to that
partons.

Now, consider a particle of mass $M$ that obtains  a boost $\psi$
from the rest-frame. Its momentum reads as,
    $$
P^\mu =  (P^+,{M \over 2P^+},0_t),
    $$
with $P^+$ being the positive light-cone momentum of the particle.
In general, one is able to write its  momentum based on the rapidity
($y$), which is defined as:
  $$
y ={1 \over 2} ln {P^+ \over P^-}.
    $$
The   result is the following expression for the momentum of the
particle:
     $$
P^\mu =(\sqrt{{M^2 +{P_t}^2}\over 2} e^y ,\sqrt{{M^2 +{P_t}^2}\over
2} e^{-y} ,P_t ),
    $$
  where $\sqrt{M^2 +{P_t}^2}$ is the so-called transverse energy of the particle, $E_t$,  \cite{light-cone}.
 Using the above method  and the conservation of energy-momentum, we can derive the following relations
 for the subprocesses $ q^* + \bar{q}^* \longrightarrow \gamma+\gamma$ and $ g^* + g^* \longrightarrow \gamma+\gamma$ :
 $$
\textbf{k}_{1,t} + \textbf{k}_{2,t} = \textbf{k}_{3,t} +
\textbf{k}_{4,t} ,
$$
    $$
x_1 = {\left( \vert k_{3,t}\vert e^{y_3} +\vert k_{4,t}\vert e^{y_4}
\right)}/\sqrt{s},
    $$
        \begin{equation}
x_2 = {\left( \vert k_{3,t}\vert e^{-y_3} +\vert k_{4,t}\vert
e^{-y_4}  \right)}/\sqrt{s}.
    \label{eq24}
    \end{equation}
 Similarly for the subprocess $g^*+q^*(\bar{q}^*)\longrightarrow
 \gamma+\gamma+q(\bar{q})$, we find:
 $$
\textbf{k}_{1,t} + \textbf{k}_{2,t} = \textbf{k}_{3,t} +
\textbf{k}_{4,t} +\textbf{k}_{5,t} ,
$$
    $$
x_1 = {\left( \vert k_{3,t}\vert e^{y_3} +\vert k_{4,t}\vert e^{y_4}
+M_{5,t} e^{y_5} \right)}/\sqrt{s},
    $$
        \begin{equation}
x_2  = {\left( \vert k_{3,t}\vert e^{-y_3} +\vert k_{4,t}\vert
e^{-y_4} +M_{5,t} e^{-y_5} \right)}/\sqrt{s}.
    \label{eq24p}
    \end{equation}
$k_{i,t}$ and $y_i$, $i=3,4$ are the transverse momentum and the
rapidity of outgoing particles, respectively. $M_{5,t}$ is the
transverse mass of produced quark or anti-quark  with mass $m$ that
is defined by:
$$M_{5,t}=\sqrt{m^2 +{\vert k_{5,t}\vert}^2},$$
while $y_5$ is its rapidity (the quarks masses are set equal to zero
as it is stated above the equation (\ref{eq005})).

\begin{figure}[H]
\includegraphics[scale=0.3]{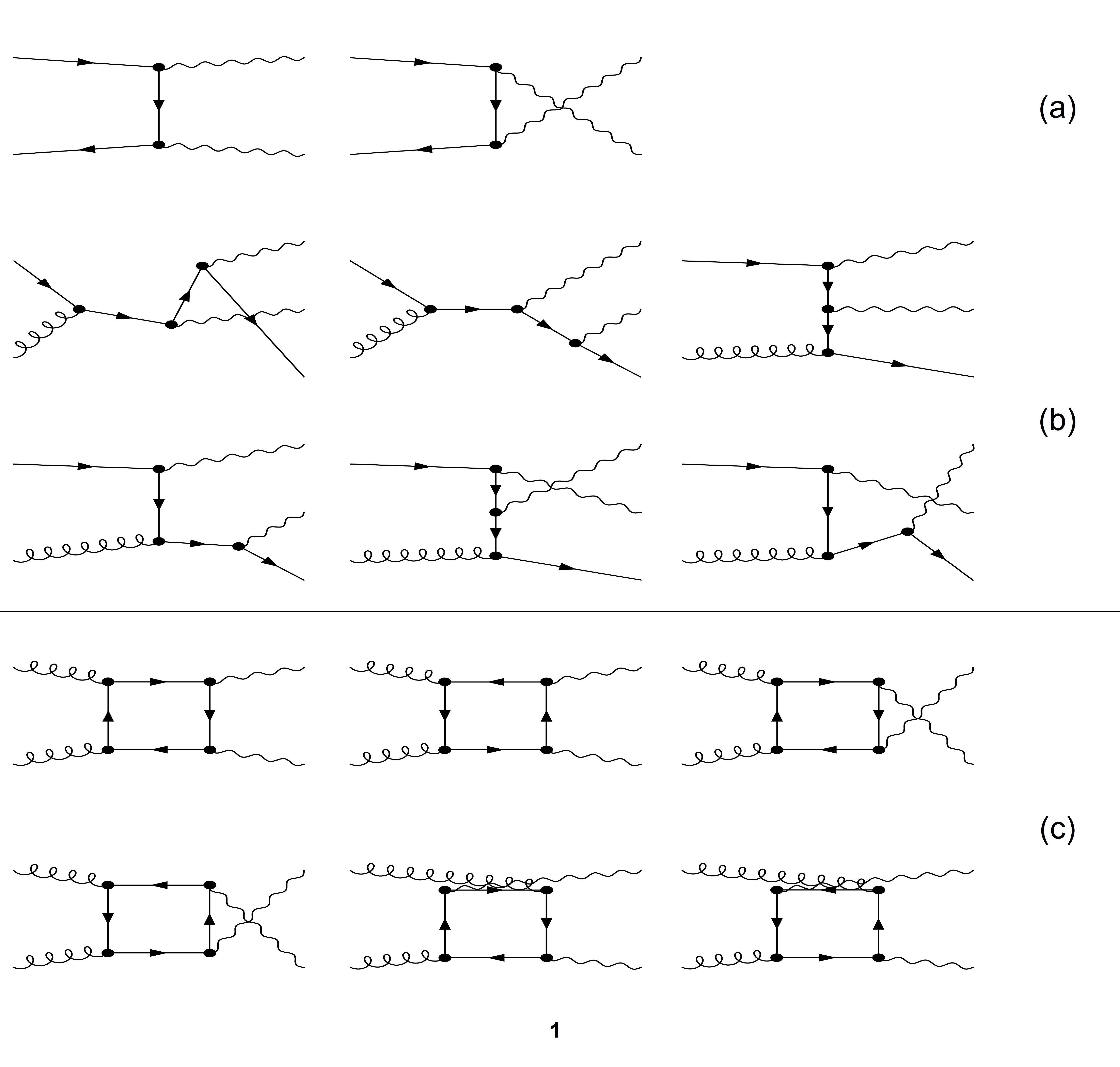}
\caption{All diagrams that are calculated for the IPPP production in
this paper. The diagrams in the panel (a) correspond to the $ q^* +
\bar{q}^* \longrightarrow \gamma+\gamma$ sub-process, the panel (b)
to the
 $g^*+q^*(\bar{q}^*)\longrightarrow \gamma+\gamma+q(\bar{q})$ sub-process and the panel (c) to the $ g^* + g^* \longrightarrow \gamma+\gamma$ sub-process.}
\label{zfig2}
\end{figure}

In this work, we consider the simplest processes for the isolated
prompt photon pair production. Therefore, the diagrams $2\rightarrow
2$ $(q^*+\bar{q}^*$ or $g^*+g^* \rightarrow \gamma+\gamma)$ and
$2\rightarrow 3$ $(q^*(\bar{q}^*)+ g^* \rightarrow
\gamma+\gamma+q(\bar{q}))$ are selected using the figure
\ref{zfig2}. The incoming legs in the figure \ref{zfig2} can be the
(anti)-quarks or gluons UPDF according to the $k_t$-factorization
procedure of corresponding differential cross section calculation.
We also investigate the dependence of the differential cross section
to the three prescriptions of UPDF (see the section III).

Furthermore,  it can be demonstrated that, the matrix element of all
diagrams for the  $ q^* + \bar{q}^* \to \gamma+\gamma$ sub-process
is as follows:
     $$
\mathcal{M} = e^2  e_q^2  \epsilon_\mu(k_3)  \epsilon_\nu(k_4)
\bar{U}(k_2)  \left( \gamma^{\nu} \; {\slashed{k_1}-\slashed{k_3}+m
\over (k_1-k_3)^2 - m^2} \; \gamma^{\mu} \; + \gamma^{\mu}
{\slashed{k_1}-\slashed{k_4}+m \over (k_1-k_4)^2 - m^2}
\;\gamma^{\nu} \right) U(k_1),
    $$
 where $e$ and $e_q$ are the electron charge and the quark electric charge respectively and  $\varepsilon_3$ and $\varepsilon_4$
 are the polarization 4-vectors of the isolated  prompt photons,  that satisfy the co-variant
 equation:
        \begin{equation}
\sum_{i=3,4} \varepsilon^{\mu}(k_i) \varepsilon^{*
\upsilon}(k_i)=-g^{\mu \upsilon}.
    \label{eq002p}
    \end{equation}
Similarly for the $g^*+q^*(\bar{q}^*)\to \gamma+\gamma+q(\bar{q})$
sub-process, the transition amplitude is:
        \begin{equation}
\mathcal{M} = e^2 \; e_q^2 \sqrt{4\pi \alpha_s} T^{\alpha}
\epsilon_\xi(k_2) \epsilon_\mu(k_3)  \epsilon_\nu(k_4) \bar{U}(k_5)
[ \sum_{i=1,6} \mathcal{A}^{\xi\mu\nu}_i ] U(k_1),\label{eq0003}
    \end{equation}
where $\mathcal{A}^{\xi\mu\nu}_i$ are defined in the reference
\cite{LIPIPPP}.

$T^\alpha$ are the generators of the $SU(3)$   color gauge group, as
the color transition operators, that are defined in the relation
with the Gell-Mann matrices ($\lambda^{\alpha}$),
           $$
T^\alpha \; = {\lambda^{\alpha}\over2}.
    $$
$\varepsilon_\xi(k_2)$ is the polarization vector of the incoming
off-shell gluon which should be modified with the eikonal vertex
(i.e the BFKL prescription, see the reference \cite{Deak1}). One
choice is to impose the so called non-sense polarization conditions
on $\varepsilon_\xi(k_2)$  which is not normalized to one
\cite{Deak1,collins} (and it will not be used in the present work):
$$
    \epsilon_{\xi}(k_2) = {2 k_{2,\xi} \over \sqrt{s}}.
    $$
But in the case of $k_t$-factorization scheme and the off-shell
gluons, the better choice is $\epsilon^\mu_2(k_2)={k^\mu_{2,t} \over
|k_{2,t}|}$, which leads to the following identity and can be easily
implemented in our calculation \cite{Deak1,collins}:
         \begin{equation}
\sum \varepsilon^{\mu}(k_2) \varepsilon^{*
\upsilon}(k_2)={{k_{2,t}^{\mu} k_{2,t}^{\upsilon}}\over k_{2,t}^2}.
    \label{eq003}
    \end{equation}
Finally, for the matrix element of the $g^*+g^*\to \gamma + \gamma $
sub-process, we use those which was calculated before by Berger et
al. {\cite{berger}},  with this difference that the kinematics given
in the equations (\ref{eq24}) and  (\ref{eq24p}) is imposed
\cite{LIPIPPP}.

So, generally the cross-section of $IPPP$ production is:
          \begin{equation}
\sigma_{\gamma \gamma}=\sum_{i,j=\;q ,g} \int
\hat{\sigma}_{ij}(x_1,x_2,\mu^2) f_i(x_1,\mu^2) f_j(x_2,\mu^2) dx_1
dx_2, \label{eq006}
    \end{equation}
 where $\hat{\sigma}_{ij}(x_1,x_2,\mu^2)$ is the partonic cross-section and $f_j(x_i,\mu^2)$ (= $xa_j(x_i,\mu^2)$) are parton
 distribution function   refer to the incoming parton $i$ that depends on two
 variables, $x$   and $\mu^2$ as the scales of the hard process.
 But in  the high-energy  domain, using the $k_t$-factorization theory, we could rewrite the
 collinear cross-section, i.e., the equation (\ref{eq006}),
 as,
        $$
\sigma_{\gamma \gamma} = \sum_{a_1,a_2=q,g} \int_0^1 {dx_1 \over
x_1} \int_0^1 {dx_2 \over x_2} \int_{0}^{\infty} {dk^2_{1,t} \over
k^2_{1,t}} \int_{0}^{\infty} {dk^2_{2,t} \over
k^2_{2,t}}\int_{0}^{2\pi} {d\phi_1 \over 2\pi}
\int_{0}^{2\pi}{d\phi_2 \over 2\pi} f_{a_1}(x_1,k^2_{1,t},\mu_1^2)
f_{a_2}(x_2,k^2_{2,t},\mu_2^2)    $$
    \begin{equation}
    \times
\hat{\sigma}_{a_1 a_2}(x_1,k^2_{1,t},\mu_1^2;x_2,k^2_{2,t},\mu_2^2),
    \label{eq007}
    \end{equation}
where $f(x_i,k_{t,1}^2,\mu^2)$ are the  $UPDF$ that depend on three
parameters, i.e. $x$, $k^2_t$ and $\mu^2$.

The UPDF are directly obtained from the PDF by using different
prescriptions (see the next section). In this paper, we  use the
three approaches namely KMR \cite{KMR}, LO-MRW and  NLO-MRW
\cite{MRW} to generate the UPDF from the PDF,  to be inserted in the
equation (\ref{eq007}).

In general, one should consider  the KMR or MRW parton densities in
the $k_t$-factorization calculations correspond to non-normalized
probability functions. They are used as the weight of the given
transition amplitudes (the matrix elements in these cases). The
transverse momentum dependence of the  UPDF comes from considering
all possible splittings up to and including the last splitting, see
the references \cite{KMR,WattWZ,WATT,KMR1}, while the evolution up
to the hard scale without change in the $k_t$, due to virtual
contributions, is encapsulated in the Sudakov-like survival form
factor. Therefore, all splittings and real emissions of the partons,
including the last emission, are factorized in the function $f_g(x,
k_t^2, \mu^2)$ as its definition.  The last emission from the
definition of the produced UPDF can not be   disassociated and to be
count    as the part of the $2\to 3$ diagrams. This point also
discussed in the reference \cite{W/Z-LHCb} and it is in contrast to
the reference \cite{LIPIPPP,LIP2011} i.e. there may not be any
double counting by taking into account $q^* + {\bar q^*} \rightarrow
\gamma + \gamma$ and $g^*+q^*({\bar q^*})
 \rightarrow q({\bar q}) + \gamma + \gamma$ processes together in the
$k_t$-factorization approach in the present calculation
\cite{KIMBER}. On the other hand some authors do believe on the
double counting. The argument goes as follows: in the region where
 the transverse momentum  of one of the parton is as large as the hard
scale and the additional parton is highly separated in the rapidity
from the hard process (multi-Regge region), the additional emission
in the 2$\rightarrow$3 should be subtracted i.e. considering the
definition of the UPDF.  However we will check the above agrement in
the sections IV and V, by modification of the UPDF to find out about
the possible double counting. One should also not that the UPDF
should satisfy the condition given in the  equation (1). So any
changes in the UPDF certainly affect the original PDF definitions.
\section{The $k_t$-factorization framework }
In the equation (\ref{eq006}) all partons are usually assumed to
move in the plane of the  incoming protons. Therefore,  they do not
posses any transverse momenta. This is so called the collinear
approximation (see the appendix A). However, at high energies and in
the small-$x$ region, the transverse momentum, $k_t$, of the
incoming partons are expected to become important. Therefore, the
cross-sections are factorized into the $k_t$-dependent partonic
cross-sections $\hat{\sigma}(x,k_t^2,\mu^2)$, where the incoming
partons are treated as the off-shell particles. So, one should use
the UPDF ($f(x, k^2_t,\mu^2)$)
   instead of the PDF in the equation (\ref{eq006}), according to
 the equation (\ref{eq0065}) which leads to the equation (\ref{eq007}).

In the rest of this section, we   briefly explained  how to evaluate
these UPPF in the simplistic frameworks.
\subsection{The KMR prescription}
The KMR UPDF are generated through a procedure that was proposed by
Kimber, Martin and Ryskin (KMR) \cite{KMR}. In this method, the UPDF
are generated  such that the partons developed from some starting
parameterizations  up to the scale $k_t$ according to the DGLAP
evolution equations.
  So the partons are evolved in the single evolution ladder (carrying only the $k_t^2$ dependency) and get convoluted
with the second scale ($\mu^2$) at the hard process. This is the
last-step evolution approximation. Then the $k_t$ is forced to
depend on the scale $\mu^2$, without any real emission, and there is
a summation  over the virtual contributions by  imposing the Sudakov
form factor ($T_a(k_t^2,\mu^2)$). So, the general form of the
KMR-UPDF are:
   \begin{equation}
f_a(x,k_t^2,\mu^2) = T_a(k_t^2,\mu^2)\sum_{b=q,g} \left[
{\alpha_S(k_t^2) \over 2\pi} \int^{1-\Delta}_{x} dz P_{ab}^{(LO)}(z)
b\left( {x \over z}, k_t^2 \right) \right] , \label{eq56}
    \end{equation}
where $T_a(k_t^2,\mu^2)$ are :
   \begin{equation}
T_a(k_t^2,\mu^2) = exp \left( - \int_{k_t^2}^{\mu^2} {\alpha_S(k^2)
\over 2\pi} {dk^{2} \over k^2} \sum_{b=q,g} \int^{1-\Delta}_{0} dz'
P_{ab}^{(LO)}(z') \right). \label{eq5}
    \end{equation}
$T_a$ are considered to be unity for $k_t>\mu$. In the above
equation $\Delta$  is proposed to prevent the soft gluon
singularity, but this constraint is imposed on the quark radiations
too.  The angular ordering constraint is imposed  to determine
$\Delta$,. Angular ordering originates from the color coherence
effects of the gluon radiations \cite{KMR}. So $\Delta$ is:
$$ \Delta = {k_t \over \mu + k_t}. $$
The $P_{ab}^{(LO)}(z)$ are the familiar LO splitting functions
\cite{collins}.
\subsection{The LO-MRW prescription}
The LO-MRW  formalism, similar to the KMR scheme,  was proposed by
Martin, Ryskin and Watt  (MRW) \cite{MRW}. This formalism has the
same general structure as the KMR, but only with one significant
difference that: the angular ordering constraint is correctly
imposed only on the on-shell radiated gluons, i.e. the diagonal
splitting functions $P_{qq}(z)$ and $P_{gg}(z)$ \cite{MRW}. So, the
LO-MRW prescription is written as:
    $$
f_q^{LO}(x,k_t^2,\mu^2)= T_q(k_t^2,\mu^2) {\alpha_S(k_t^2) \over
2\pi} \int_x^1 dz \left[ P_{qq}^{(LO)}(z) {x \over z} q \left( {x
\over z} , k_t^2 \right) \Theta \left( {\mu \over \mu + k_t}-z
\right) \right.
    $$
    \begin{equation}
\left. + P_{qg}^{(LO)}(z) {x \over z} g \left( {x \over z} , k_t^2
\right) \right], \label{eq7}
    \end{equation}
with
    \begin{equation}
T_q(k_t^2,\mu^2) = exp \left( - \int_{k_t^2}^{\mu^2} {\alpha_S(k^2)
\over 2\pi} {dk^{2} \over k^2} \int^{z_{max}}_{0} dz'
P_{qq}^{(LO)}(z') \right), \label{eq8}    \end{equation} for the
quarks and
    $$
f_g^{LO}(x,k_t^2,\mu^2)= T_g(k_t^2,\mu^2) {\alpha_S(k_t^2) \over
2\pi} \int_x^1 dz \left[ P_{gq}^{(LO)}(z) \sum_q {x \over z} q
\left( {x \over z} , k_t^2 \right)
    \right.$$
    \begin{equation}
\left. + P_{gg}^{(LO)}(z) {x \over z} g \left( {x \over z} , k_t^2
\right) \Theta \left( {\mu \over \mu + k_t}-z \right)
    \right], \label{eq9}
    \end{equation}
with
    \begin{equation}
T_g(k_t^2,\mu^2) = exp \left( - \int_{k_t^2}^{\mu^2} {\alpha_S(k^2)
\over 2\pi} {dk^{2} \over k^2}
    \left[ \int^{z_{max}}_{z_{min}} dz' z' P_{qq}^{(LO)}(z')
+ n_f \int^1_0 dz' P_{qg}^{(LO)}(z') \right] \right) , \label{eq10}
    \end{equation}
for the gluons. In the equations (\ref{eq8}) and (\ref{eq10}),
$z_{max}=1-z_{min}=\mu/(\mu+k_t)$ \cite{WATT}. The UPDF of KMR and
MRW to a good approximation, include the main kinematical effects
involved in the  IS processes.  Note that the particular choice of
the AOC in the KMR formalism despite being of the LO, includes some
contributions from the NLO sector, hence in the case of MRW
framework, these contributions must be inserted separately.
\subsection{The NLO-MRW prescription}
Finally,  MRW  \cite{MRW}  proposed a method for the promotion of
the LO-MRW to the NLO-MRW prescription. Utilizing the NLO PDF and
corresponding splitting functions from DGLAP evolution equations
lead to the MRW-NLO formalism \cite{MRW}. The general form of the
NLO-MRW UPDF are:
    $$
f_a^{NLO}(x,k_t^2,\mu^2)= \int_x^1 dz T_a \left( k^2={k_t^2 \over
(1-z)}, \mu^2 \right) {\alpha_S(k^2) \over 2\pi}
    \sum_{b=q,g} \tilde{P}_{ab}^{(LO+NLO)}(z)
    $$
    \begin{equation}
\times b^{NLO} \left( {x \over z} , k^2 \right) \Theta \left(
1-z-{k_t^2 \over \mu^2} \right),
    \label{eq11}
    \end{equation}
with the "extended" NLO splitting functions,
$\tilde{P}_{ab}^{(i)}(z)$,  being defined as,
    \begin{equation}
\tilde{P}_{ab}^{(LO+NLO)}(z) = \tilde{P}_{ab}^{(LO)}(z) + {\alpha_S
\over 2\pi}
    \tilde{P}_{ab}^{(NLO)}(z),
    \label{eq12}
    \end{equation}
and
    \begin{equation}
\tilde{P}_{ab}^{(i)}(z) = P_{ab}^{i}(z) - \Theta (z-(1-\Delta))
\delta_{ab} F^{i}_{ab} P_{ab}(z),
    \label{eq13}
    \end{equation}
where $i= 0$ and $1$ stand for the LO and the NLO, respectively. The
reader can find a comprehensive description of the NLO splitting
functions in the references \cite{MRW,PNLO}. We must however
emphasize   that  in contrary to the KMR and the LO-MRW frameworks,
the AOC is being introduced into the NLO-MRW formalism via the
$\Theta (z-(1-\Delta))$ constraint, in the "extended" splitting
function. Now $\Delta$ can be defined as:
    $$ \Delta = {k\sqrt{1-z} \over k\sqrt{1-z} + \mu}.$$
 This framework are  the
collection of the NLO PDF, the NLO splitting functions and the
constraint $\Theta \left( 1-z-{k_t^2 / \mu^2} \right)$ which impose
the NLO corrections to this method. Nevertheless, it was shown that
using only the LO part of the "extended" splitting functions,
instead of the complete definition of the equation (\ref{eq12}),
would result a reasonable accuracy in the computation of the NLO MRW
UPDF \cite{MRW}. Additionally, the Sudakov form factors in this
framework are defined as:
    \begin{equation}
T_q(k^2,\mu^2) = exp \left( - \int_{k^2}^{\mu^2} {\alpha_S(q^2)
\over 2\pi} {dq^{2} \over q^2} \int^1_0 dz' z' \left[
\tilde{P}_{qq}^{(0+1)}(z') + \tilde{P}_{gq}^{(0+1)}(z') \right]
\right) ,
    \label{eq14}
    \end{equation}
    \begin{equation}
T_g(k^2,\mu^2) = exp \left( - \int_{k^2}^{\mu^2} {\alpha_S(q^2)
\over 2\pi} {dq^{2} \over q^2} \int^1_0 dz' z' \left[
\tilde{P}_{gg}^{(0+1)}(z') + 2n_f\tilde{P}_{qg}^{(0+1)}(z') \right]
\right) .
    \label{eq15}
    \end{equation}
Each of  the KMR, the LO and the NLO MRW  UPDF can be used to
identify the probability of finding a parton of a given flavor, with
the fraction $x$ of longitudinal momentum of the parent hadron and
the transverse momentum $k_t$, in the scale $\mu^2$ at the semi-hard
level of a particular IS process.

The  modifications to the above KMR, LO-MRW and NLO-MRW UPDF are
made in our calculation of IPPP production cross sections, in the
section V, to investigate the possible double counting concerning
the $2\rightarrow 3$ process, according to the reference
\cite{LIP2011}
\section{the IPPP production and the technical prescription}
 For calculating the partonic cross-section, we need   the  matrix element squared ($|\mathcal{M}|^2 $)
 of sub-processes. Since the incoming quarks and  gluons are off-shell, the expression
 for such matrix element will be more complicated. Therefore, we use the BFKL
 prescription \cite{Deak1} for the gluons in the equation (\ref{eq003}) and apply the method proposed
  in the references \cite{lip31,lip32} for the incoming quarks, for the small x region. In this method, it is assumed that the incoming
  quarks with 4-momenta ($p_\mu$) radiate a gluon (or a photon) and consequently become off-shell \cite{LIPIPPP,W/Z-NLO}.
  Therefore the extended $|\mathcal{M}|^2$ becomes,
           \begin{equation}
|\mathcal{M}|^2 \sim | \tilde{T}^{\alpha}  {\slashed{k}+m \over
(k)^2 - m^2} \gamma_{\beta} \; U(p) \bar{U   }(p) \; \gamma^{\beta}
{\slashed{k}+m \over (k)^2 - m^2} T_{\alpha}  |,
    \label{eq004p}
    \end{equation}
    where $\tilde{T}^\alpha$ and $T_\alpha$ represent the rest of the matrix elements.
 Since, the expression presented between $T_{\alpha}$
and $\tilde{T}^\alpha$ is considered to be the off-shell quark spin
density matrix elements, then by  using the on-shell identity,
performing some Dirac algebra at the $m \to 0$ limit and imposing
the Sudakov decomposition, $k = xp + k_t$, with $k^2 = k^2_t =
-\textbf{k}^2_t$ and some straightforward algebra  we obtain
\cite{W/Z-NLO,Deak1,LIPIPPP}:
\begin{equation}
|\mathcal{M}|^2 \sim {{ 2\over {xk^2}} tr[ T^{\mu} x \hat{p}
\tilde{T}_\mu ] }. \label{eq005}
    \end{equation}
where $x \hat{p}$ represents the properly normalized off-shell spin
density matrix.

Since the calculation of the $|\mathcal{M}|^2$ is a laborious task,
we use the algebraic manipulation system $\mathtt{FORM}$
\cite{FORM}. The above approximations, which are valid at small x
region, force some limits on our kinematics range. So the resulted
differential cross section may not cover the whole experimental data
of Tevatron and LHC colliders (see our discussion in the section V)
\cite{LIPIPPP,KUTAK1,KUTAK2,KUTAK3,KUTAK4,LIP2016}.

In the section II, we defined the total cross-section for the IPPP
production at   hadronic collisions, $\sigma_{\gamma\gamma}$, as:
    \begin{equation}
d\hat{\sigma}_{a_1 a_2} = {d\phi_{a_1 a_2} \over F_{a_1 a_2}}
|{\mathcal{M}_{a_1 a_2}}|^2.
    \label{eq17}
    \end{equation}
$d\phi_{a_1 a_2}$ and $F_{a_1 a_2}$ are the multi-particle phase
apace and the flux factor, respectively which can be defined
according to the specifications of the partonic process,
    \begin{equation}
d\phi_{a_1 a_2} = \prod_i {d^3 p_i \over 2E_i} \delta^{(4)}\left(
\sum p_{in} - \sum p_{out} \right) ,
    \label{eq18}
    \end{equation}
    \begin{equation}
    F_{a_1 a_2} = x_1 x_2 s,
    \label{eq19}
    \end{equation}
where the $s$ is the center of mass energy squared,
    $$ s=(P_1 + P_2)^2=2P_1.P_2. $$
$d\phi_{a_1 a_2}$ can be characterized in terms of the transverse
momenta of the product particles $p_{i,t}$, their rapidities, $y_i$,
and the azimuthal angles of the emissions, $\varphi_i$,
    \begin{equation}
{d^3 p_i \over 2E_i} = {\pi \over 2} dp_{i,t}^2 dy_i {d\varphi_i
\over 2\pi}.
    \label{eq20}
    \end{equation}
In the present work, ${\mathcal{M}_{a_1 a_2}}$ in the equation
(\ref{eq17}),  are the matrix elements of the partonic diagrams
which are involved in the production of the final results (see the
section II).

By using the kinematics given in the     section II, we can derive
the following equations for the total cross-section of the IPPP
production in the framework of $k_t$-factorization. So the total
cross-section for $q \bar{q}$ and $gg$  are:
$$
\sigma (P+\bar{P} \rightarrow\gamma+\gamma) = \sum_{a_i,b_i = q,g}
\int {dk_{a_1,t}^2 \over k_{a_1,t}^2} \; {dk_{a_2,t}^2 \over
k_{a_2,t}^2} \;
    dp_{\gamma1,t}^2 \;  dy_{\gamma1} \; dy_{\gamma2} \; {d\varphi_{a_1} \over 2\pi} \; {d\varphi_{a_2}\over 2\pi} \;{d\varphi_{\gamma_1}\over 2\pi} \;
    \; \times
    $$
      \begin{equation}
{|\mathcal{M}(a_1+a_2 \rightarrow \gamma+\gamma)|^2 \over 16 \pi
(x_1 x_2 s)^2} \; f_{a_1}(x_1,k_{a_1,t}^2,\mu^2) \;
f_{a_2}(x_2,k_{a_2,t}^2,\mu^2),
    \label{eq251}
    \end{equation}
and for  $qg$ and $\bar{q}g$ are,
 $$
\sigma (P+\bar{P} \rightarrow\gamma+\gamma + X) = \sum_{a_i,b_i =
q,g} \int {dk_{a_1,t}^2 \over k_{a_1,t}^2} \; {dk_{a_2,t}^2 \over
k_{a_2,t}^2} \;
    dp_{\gamma_1,t}^2 \; dp_{\gamma_2,t}^2 \; dy_{\gamma1} \; dy_{\gamma2} \; dy_5
    \; \times
    $$
    $$
{d\varphi_{a_1} \over 2\pi} \; {d\varphi_{a_2} \over 2\pi} \;
{d\varphi_{\gamma_1} \over 2\pi} \; {d\varphi_{\gamma_2} \over 2\pi}
    \times
    $$
    \begin{equation}
{|\mathcal{M}(a_1+a_2 \rightarrow\gamma+\gamma+X)|^2 \over 256 \pi^3
(x_1 x_2 s)^2} \; f_{a_1}(x_1,k_{a_1,t}^2,\mu^2) \;
f_{a_2}(x_2,k_{a_2,t}^2,\mu^2),
    \label{eq261}
    \end{equation}
Note that the integration boundaries for ${dk_{i,t}^2 / k_{i,t}^2}$
are limited by the kinematics. So one can introduce an upper limit
for these integration, say $k_{i,max}$, several times larger than
the scale $\mu$. In addition, $k_{t,min}=\mu_0\sim1$ $ GeV$,  is
considered as the lower limit, that separates the non-perturbative
and the perturbative regions, by assuming that,
    \begin{equation}
{1\over {k_t^2}}\; f_a(x,k_t^2,\mu^2) {\huge |}_{k_t<\mu_0}  =
{1\over \mu_0^2}\;a(x,\mu_0^2)\;T_a(\mu_0^2,\mu^2).
    \label{eq0025}
    \end{equation}
 As a result of above formulation, the densities of partons are constant for $k_t<\mu_0$ at fix $x$ and $\mu$ \cite{MRW}.
 For the above calculations, we  use the LO-MMHT2014 PDF libraries
  for the KMR and the LO-MRW UPDF schemes, and the NLO-MMHT2014 PDF libraries for
 the  NLO-MRW formalism.

 The  VEGAS algorithm is considered for performing  the
multidimensional integration
   of the  total cross-section in the  equations (\ref{eq251}) and (\ref{eq261}).
   Since the sea quarks
  become significant in the high energy limit, we calculate  the cross-section
  of IPPP production, by considering four flavors (i.e. the up, down, charm and
  strange flavors) for CM energy of 1.960 TeV (D0 and CDF) and add bottom
   flavor for CM energy of  7 TeV (CMS and ATLAS).
 Before we present our results, it is important to have  the relations
 between the different channel parameters, i.e. $M_{\gamma\gamma}$,  $p_{t,\gamma\gamma}$,
 $cos\theta^*_{\gamma\gamma}$, $\Delta\phi_{\gamma\gamma}$,
 $z_{\gamma\gamma}$
and $y_{\gamma\gamma}$, which is given in the references
 \cite{D010,D013,CDF11,CDF13,CMS11,CMS14,ATLAS10,ATLAS13}.

Some divergences  appear because of the small $k_t$ ($<<\mu$)  of
the outgoing quark in the case of $q^*(\bar{q}^*)+g^*\rightarrow
\gamma+\gamma+q(\bar{q})$ process. But since this quark is in the
direction of outgoing photons it is eliminated by excluding the
above mentioned regions in our calculation and also implementing
isolated and separated cone in this computation. In this work, we
applied the same method as the reference \cite{DIPHOX} and
\cite{LIPIPPP}
 for the phase space cut. To avoid the
double counting and divergence, the invariant mass of the
photon-quark subsystem is considered to be greater than 1 GeV. In
order to be insure about the possible double counting in UPDF
\cite{9B,10B}, we checked our result by modifying the UPDF according
to reference \cite{LIPIPPP,LIP2011} and suppressing the quarks
splitting in the UPDF that will be discussed in the section V.
Otherwise one should perform substraction procedure
\cite{saleev1,saleev}. We should point out here that the
fragmentation contribution of the $q(\bar{q})\rightarrow \gamma$ or
$g\rightarrow \gamma$ to the $2\rightarrow 3$ processes can be
dramatically reduced by applying the same photon isolation and
separation cone implemented by the experimental setup
\cite{D010,D013,CDF11,CDF13,CMS11,CMS14,ATLAS10,ATLAS13}. Beside
this restriction, as we stated before, we also choose the invariant
mass of quark and photon subsystem to be greater than 1 GeV , in
order to eliminate any divergence from our cross section calculation
\cite{DIPHOX,LIPIPPP}.

The strong coupling  is $\Lambda_{QCD}$=200 MeV and $\alpha_s$ is
chosen to be one and two loops in case of LO and NLO level,
respectively \cite{collins}. As it was pointed out before, the same
photon isolation cuts are implemented as the one imposed in the
related experiment
\cite{D010,D013,CDF11,CDF13,CMS11,CMS14,ATLAS10,ATLAS13}. We should
mention that, the factorization scale $\mu$ is chosen such that, the
renormalization scale $\mu_{R}$ to be equal to the invariant mass of
photon-photon sub-system $ M_{\gamma\gamma}$.
\section{Results and discussions } In this section, we present
our results, regarding  the IPPP  production   according to the
experimental specifications  discussed in the appendix A. Note that
the fragmentation effects enhanced (suppressed) when
$p_{t,\gamma\gamma}>M_{\gamma\gamma}$ or
$\Delta\Phi_{\gamma\gamma}<\pi/2$ or
$\mid\cos\theta^*_{\gamma\gamma}\mid>0.6$
($p_{t,\gamma\gamma}<M_{\gamma\gamma}$ or
$\Delta\Phi_{\gamma\gamma}>\pi/2$ or
$\mid\cos\theta^*_{\gamma\gamma}\mid<0.6$). Since our calculations
show different behavior corresponding to the different CM energies,
our results and discussions are given into two subsections:
\subsection{$E_{CM}=1.96\;TeV$}
\begin{figure}[H]
\includegraphics[scale=0.3]{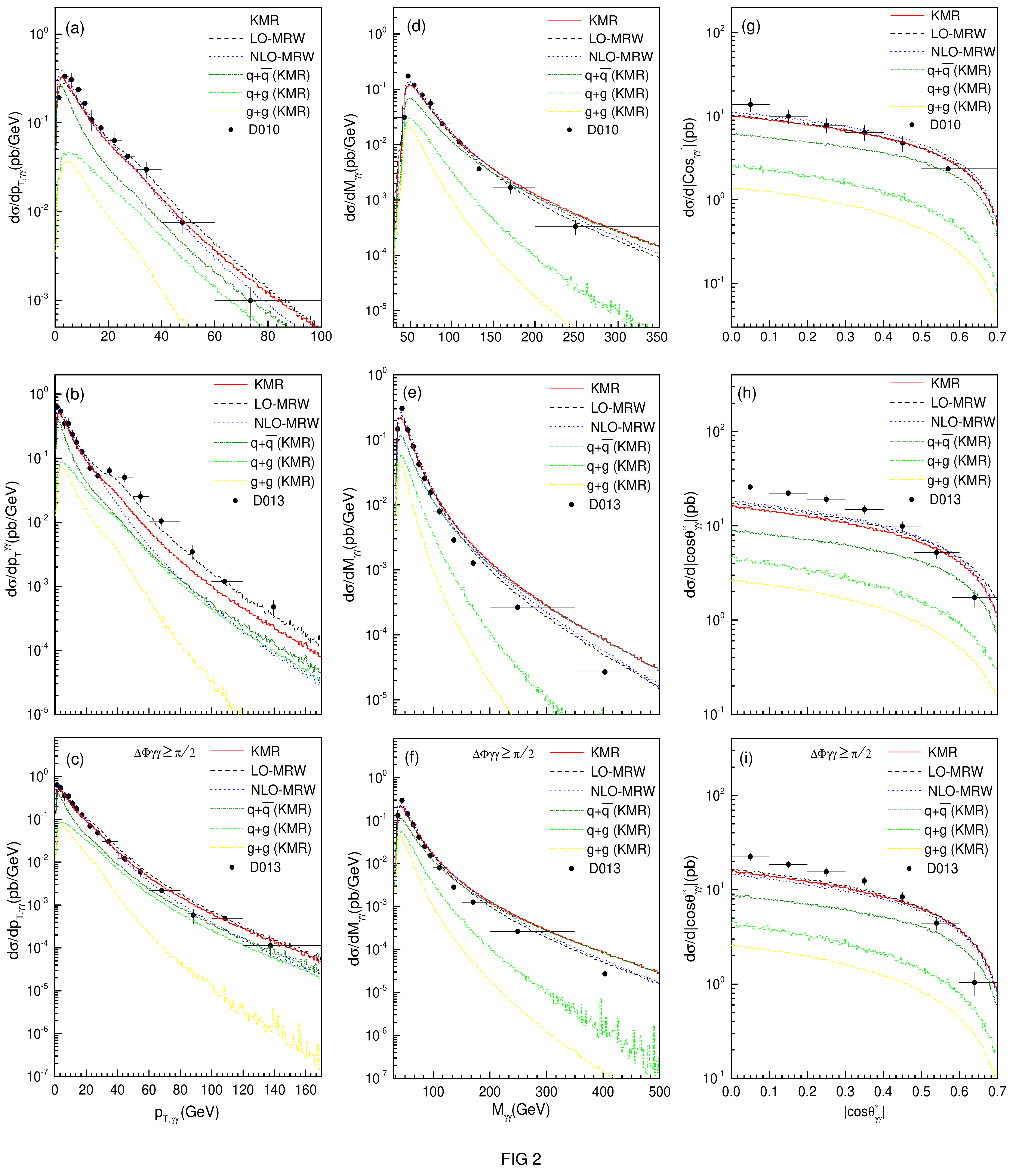}
\caption{The differential cross section of the production of $IPPP$
as functions of the transverse momentum ($p_{t,\gamma\gamma}$ in the
panels (a), (b) and (c)), the photon invariant mass
($M_{\gamma\gamma}$ in the panels (d), (e) and (f)) and
$cos\theta_{\gamma\gamma}^{*}$ (in the panels (g), (h) and (i)) at
$E_{CM}=1960 \;GeV$. The experimental data are from the $D0$
collaboration, \cite{D010,D013}. Note that the sub-processes are
only given for the KMR approach. }
 \label{fige3}
\end{figure}

\begin{figure}[H]
\includegraphics[scale=0.3]{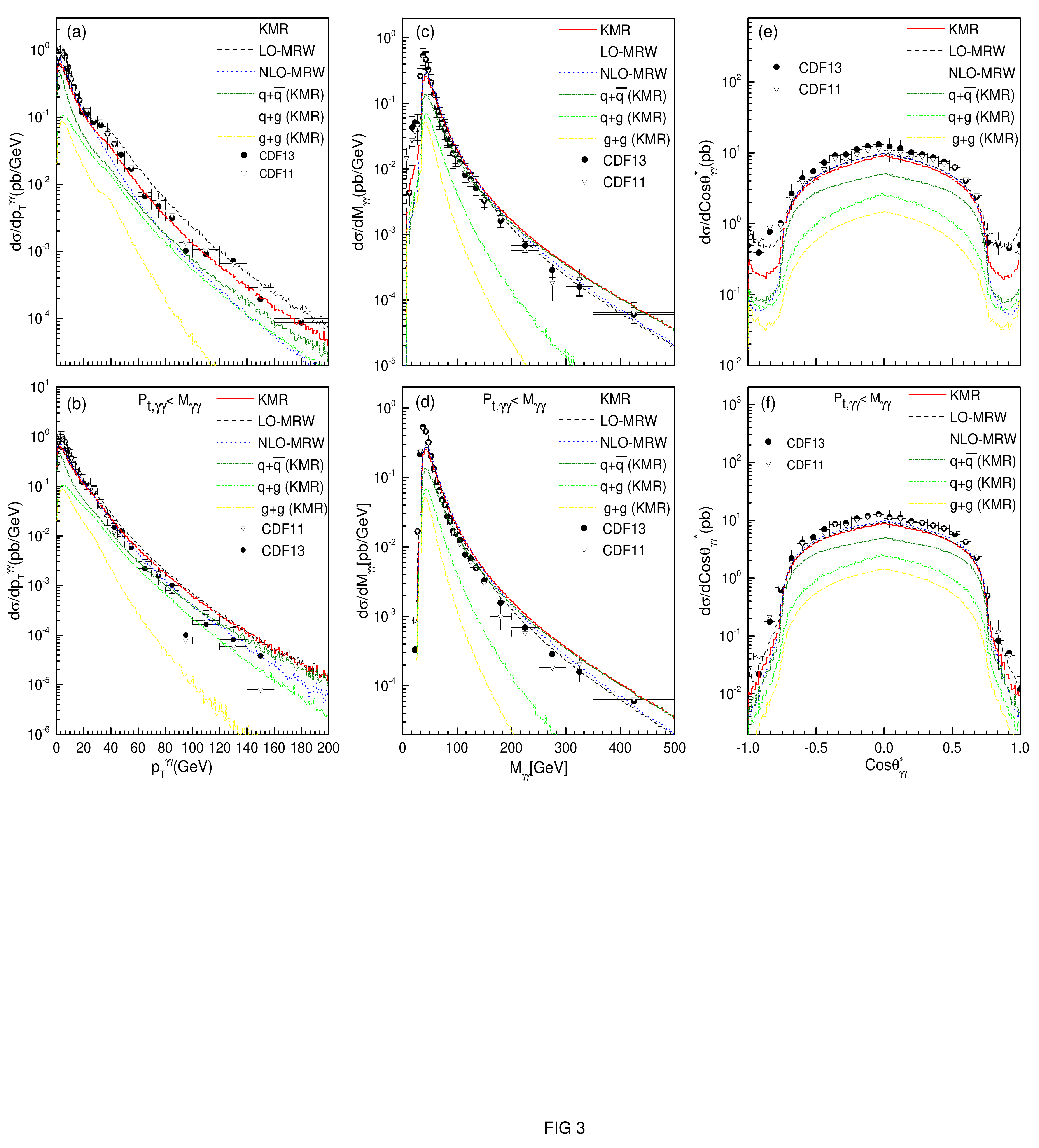}
\caption{The differential cross section of the production of $IPPP$
as functions of the transverse momentum ($p_{t,\gamma\gamma}$ in the
panels (a) and (b) ), photon invariant mass ($M_{\gamma\gamma}$ in
the panels (c) and (d)) and $cos\theta_{\gamma\gamma}^{*}$ (in the
panels (e) and (f)) at $E_{CM}=1960 \;GeV$. The experimental data
are from the  CDF  collaboration \cite{CDF11,CDF13}. Note that the
sub-processes are only given for the  KMR  approach.} \label{fige4}
\end{figure}

The figures \ref{fige3} and \ref{fige4},  illustrate our
calculations regarding the IPPP production differential cross
section in the $E_{CM}=1.96\;TeV$, in accordance to the experimental
data of the D0 (D010 and DO13) and CDF (CDF11 and CDF13)
collaborations \cite{D010,D013,CDF11,CDF13},
 using   the KMR, LO-MRW and NLO-MRW prescriptions,
 as a function of $p_{t,\gamma\gamma}$, $M_{\gamma\gamma}$ and $cos\theta^{*}_{\gamma\gamma}$, respectively.
 Note that the contribution
of the individual sub-processes i.e. $q^*+\bar{q}^* \to \gamma +
\gamma$, $q^*(\bar{q}^*)+g^* \to \gamma + \gamma+ q(\bar{q})$ and
$g^*+g^* \to \gamma + \gamma$ are given only for the $KMR$ approach.

Considering the different panels of above figures, one readily finds
 that the $q^*+\bar{q}^* \to \gamma + \gamma$ contributions dominate.
But  for larger transverse momenta ($p_{t,\gamma\gamma}$) region,
the effects of $q^*(\bar{q}^*)+g^* \to \gamma + \gamma+q(\bar{q})$
sub-process become non-negligible. Interestingly, one can see the
so-called Guillet shoulder \cite{Guilet binoth} (note that a new
channel opens beyond the leading order (NLO, NNLO etc) where the
transverse momentum of pair photons is close to the pair photons
transverse momentum cut (threshold) which makes this shoulder) is
forming in the intermediate transverse momentum range (see the
panels (a)-(c) of figure \ref{fige3} and the panels (a)-(b) of
figure \ref{fige4}). Additionally, since in the D013
 report \cite{D013}, the fragmentation effects are not fully
suppressed, this "shoulder" can be seen more clearly in the panel
(b) of the  figure \ref{fige3}. Such behavior can be seen in all of
the regions where the fragmentation and the higher order (pQCD)
effects become important.

On the other hand, in the panel (c) of the figure \ref{fige4}, one
can obviously see the "low-tail of the mass" (i.e. the small
$M_{\gamma\gamma}$ region,    where a raise in the differential
cross section is acquired), appearing at the
small-$M_{\gamma\gamma}$ region, which  strongly sensitive to the
choice of   the mid-$p_{t,\gamma\gamma}$ and the
low-$\Delta\phi_{\gamma\gamma}$ domain in the range of our and the
others  calculations \cite{NNLO}. In the panel  (e) of this figure,
 the reader should notice that in the
$|cos\theta^{*}_{\gamma\gamma}|>0.6$ region, the fragmentation
effects become important. However, such behavior is missing or
negligible in the panel (f), since the fragmentation effects are
being suppressed in these areas by the means of introducing the
$p_{t,\gamma\gamma}<M_{\gamma\gamma}$ constraint \cite{CDF11,CDF13}.

Because of the small x approximation which was made in the section
IV for the partial insurance of the gauge invariance of the partonic
cross section within high-energy factorization, our result may not
be accurate for large $M_{\gamma\gamma}$ as far as we are working in
the small x region in which the incoming partons have large
transverse momenta.

\begin{figure}[H]
\includegraphics[scale=0.3]{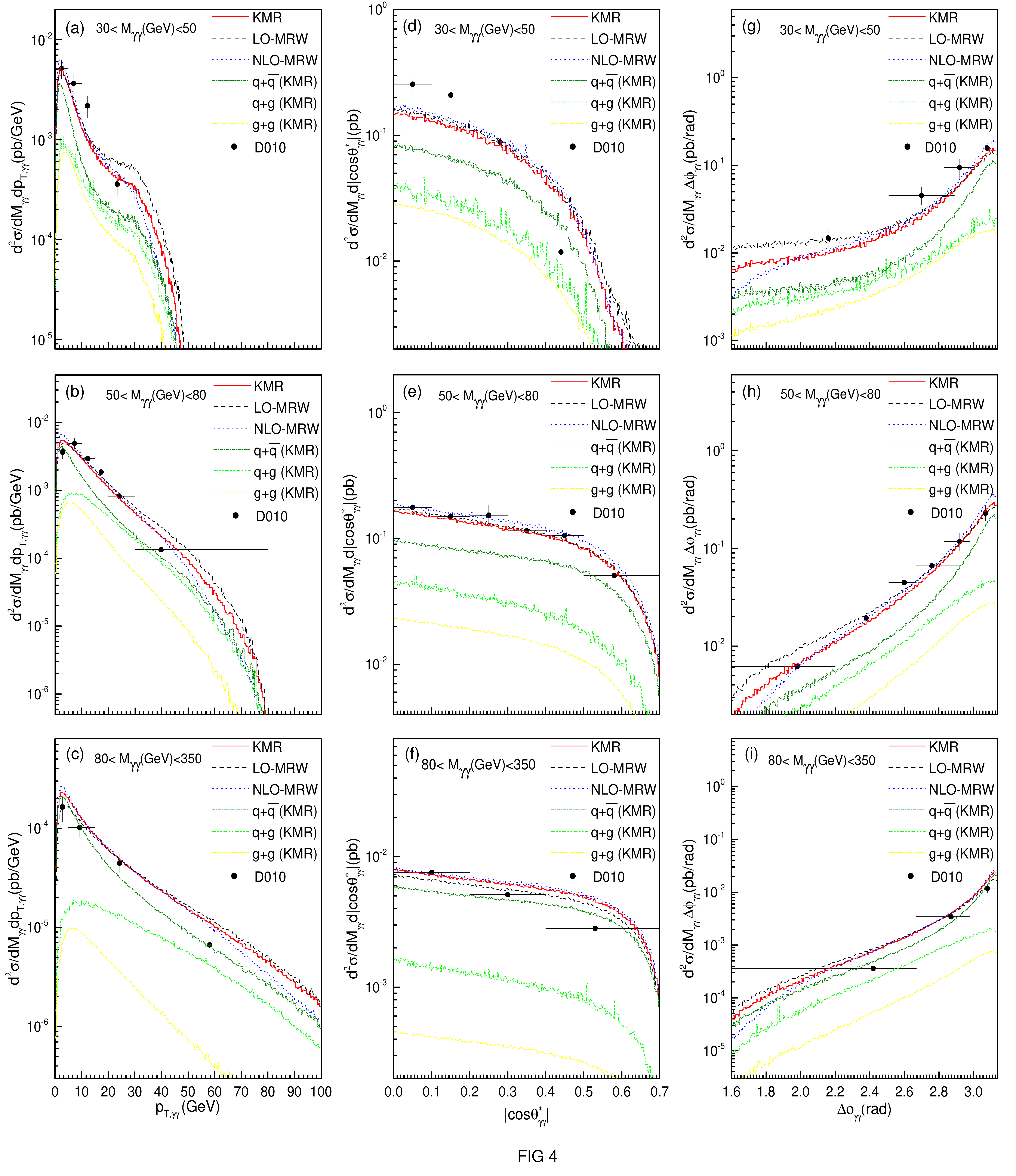}
\caption{The double-differential cross section of the production of
$IPPP$ as functions of the transverse momentum
($p_{t,\gamma\gamma}$) and $M_{\gamma\gamma}$ in the panels (a), (b)
and (c), $cos\theta_{\gamma\gamma}^{*}$ and $M_{\gamma\gamma}$ (in
the panels (d), (e) and (f)) and $ \Delta\phi_{\gamma\gamma}$ and
($M_{\gamma\gamma}$ (in the panels (g), (h) and (i)) at $E_{CM}=1960
\;GeV$. The experimental data are from the $D0$ collaboration,
\cite{D010}. Note that the sub-processes are only given for the
$KMR$ approach.} \label{fige5}
\end{figure}

Similar comparisons are made regarding the double differential cross
sections $d^2 \sigma/dp_{t,\gamma\gamma} dM_{\gamma\gamma} $, $ d^2
\sigma/\Delta\phi_{\gamma\gamma} dM_{\gamma\gamma} $ and  $ d^2
\sigma/\cos\theta^*_{\gamma\gamma} dM_{\gamma\gamma} $, in the
figure \ref{fige5}, against the data of the D010 collaboration
\cite{D010}. To be specific, what makes the difference in these
calculations, is different cuts on the $M_{\gamma\gamma}$, which
defers from $30\;GeV<M_{\gamma\gamma}<50\;GeV$,
$50\;GeV<M_{\gamma\gamma}<80\;GeV$ and
$80\;GeV<M_{\gamma\gamma}<350\;GeV$ in the figure \ref{fige5} and
they are coated  in each panel. One notices that, the best
predictions are being obtained in the
$50\;GeV<M_{\gamma\gamma}<80\;GeV$ range. Since it corresponds to
the intermediate transverse momentum regions, where (in the absence
of strong fragmentation effects) we expect to achieve the best
outcome. In the $80\;GeV<M_{\gamma\gamma}<350\;GeV$ range, the
higher order pQCD effects are larger, hence our results are
generally lower than the data.

\begin{figure}[H]
\includegraphics[scale=0.3]{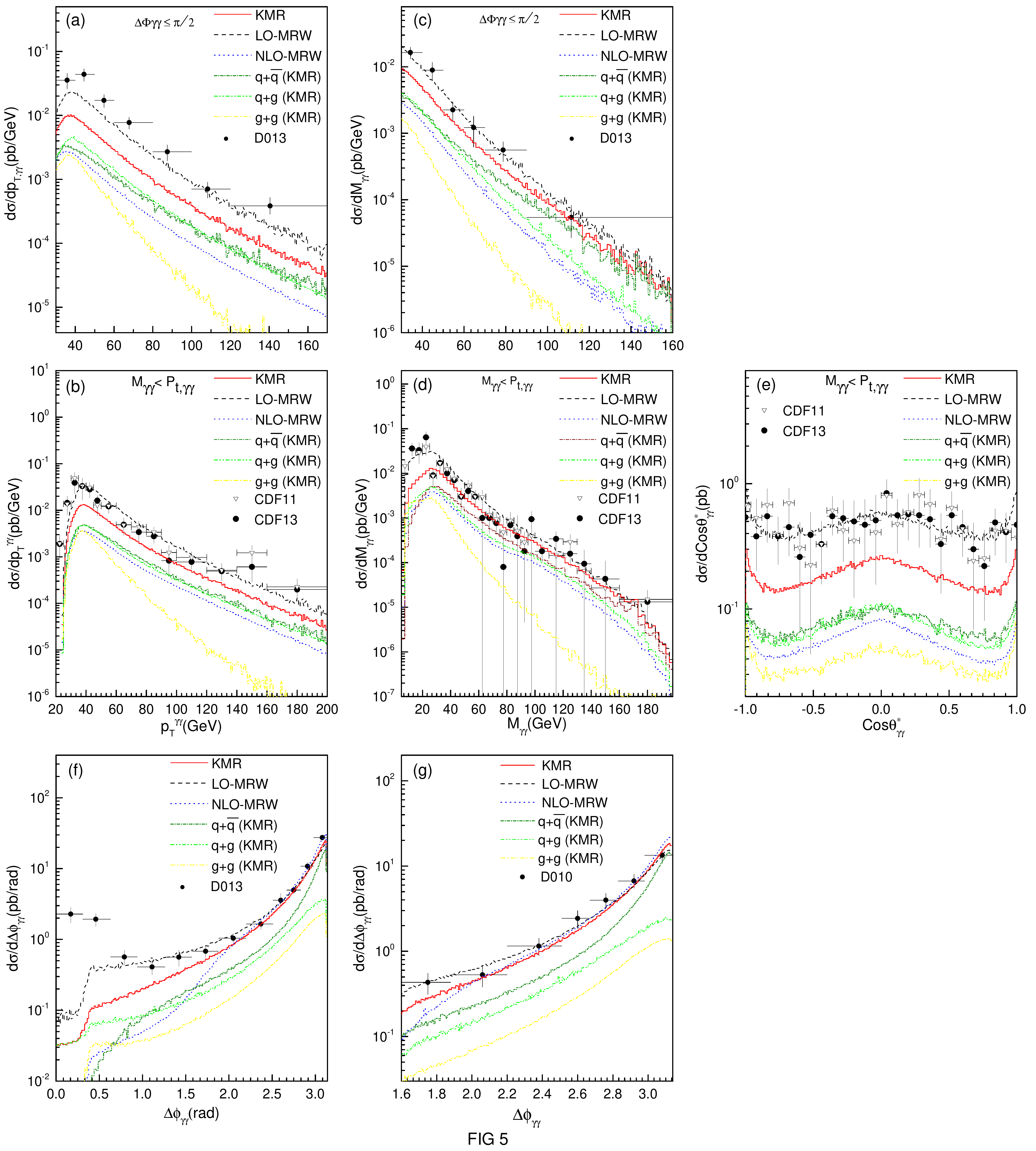}
\caption{The
 differential cross section of the production of
$IPPP$ as functions of the transverse momentum
($p_{t,\gamma\gamma}$) in the panels (a) and (b)), photon invariant
mass ($M_{\gamma\gamma}$ in panels (c) and  (d)),
$cos\theta_{\gamma\gamma}^{*}$ (in the panel (e)) and $
\Delta\phi_{\gamma\gamma}$ (in the panels (f) and (g)) at
$E_{CM}=1960 \;GeV$. The experimental data are from the $D0$ and
 CDF  collaborations, \cite{D010,D013,CDF11,CDF13}. Note that the
sub-processes are only given for the KMR approach. } \label{fige6}
\end{figure}

\begin{figure}[H]
\includegraphics[scale=0.3]{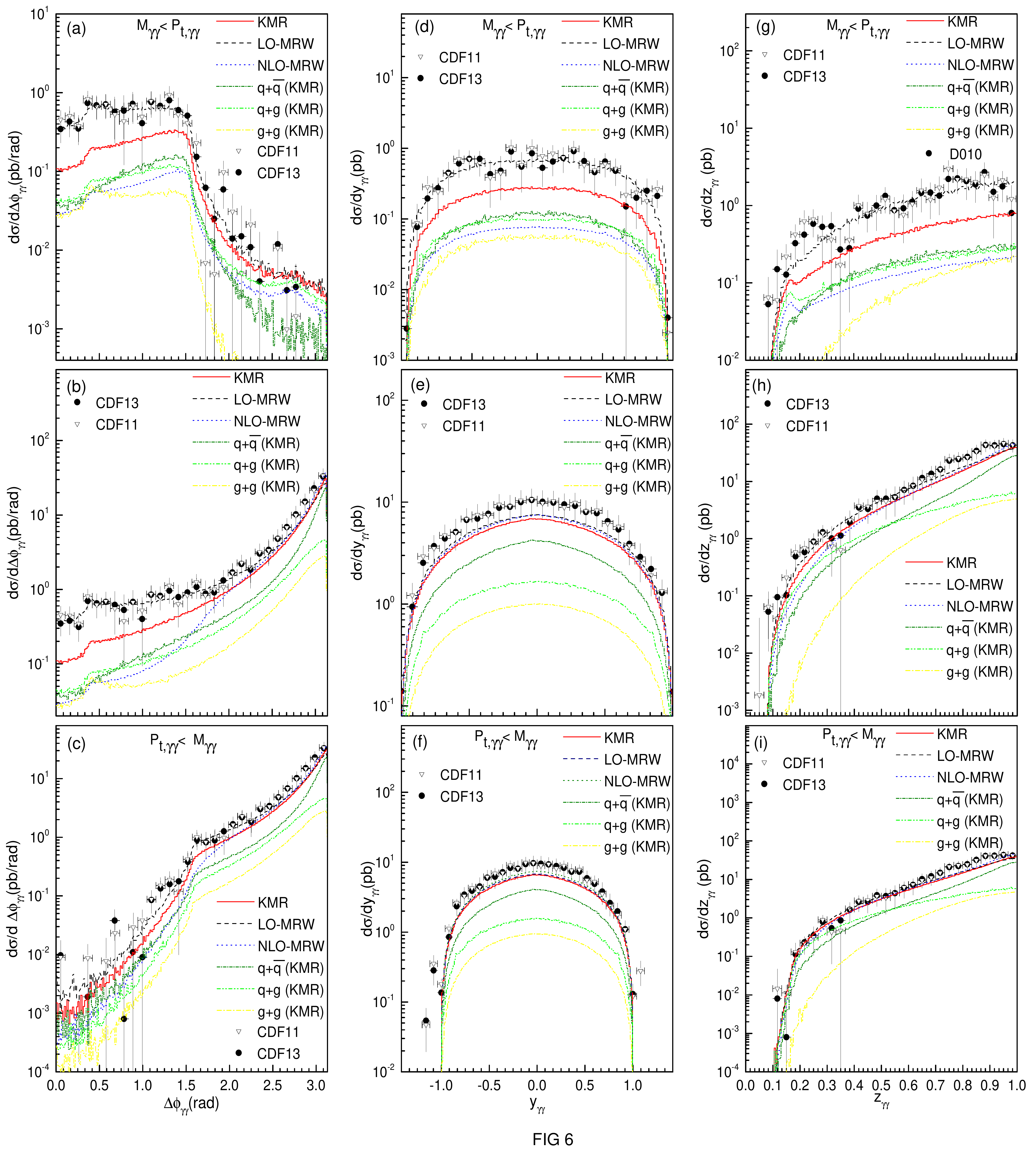}
\caption{The differential cross section of the production of $IPPP$
as functions of the ($\phi_{\gamma\gamma}$ in the panels (a), (b)
and (c)), ($y_{\gamma\gamma}$ in the panels (d), (e) and (f)) and
$Z_{\gamma\gamma}$ (in the panels (g), (h) and (i)) at $E_{CM}=1960
\;GeV$. The experimental data are from the $CDF$ collaboration,
\cite{CDF11,CDF13}. Note that the sub-processes are only given for
the KMR approach.} \label{fige7}
\end{figure}

At the $\Delta\Phi_{\gamma \gamma}\leq \pi/2$ or
$p_{t,\gamma\gamma}>M_{\gamma\gamma}$ domain, the contribution of
the fragmentation becomes utterly non-negligible, as it can be seen
in the panels (a)-(e) of the figure \ref{fige6}, as well as in the
panels (a), (d) and (g) of the figures \ref{fige7}, where the
predictions of our  simplistic framework are clearly insufficient to
describe the experimental data from the D0 and CDF collaborations
\cite{D013,CDF11,CDF13} (the constraints are presented on each
panel). To account for the missing contributions, one has to
incorporate the fragmentation and the higher-order pQCD corrections
into the our framework. Moreover, in the figure \ref{fige7}, the
reader can also find the IPPP production rates as the functions of
$y_{\gamma \gamma}$ and $z_{\gamma \gamma}$ parameters. The
symmetric form of the $y_{\gamma \gamma}$ distributions are due to
the fact that the $q^*(\bar{q}*)+g^* \to \gamma + \gamma +
q(\bar{q})$  and $g^*+q^*(\bar{q}^*) \to \gamma + \gamma +
q(\bar{q})$ are the $2 \to 3$ asymmetric processes, so their
summation becomes symmetric around $y_{\gamma \gamma}=0$. Also, the
$q^*(\bar{q}^*)+g \to \gamma + \gamma + q(\bar{q})$ sub-process
causes an interesting effect on the $z_{\gamma \gamma}$ parameter
distributions, by adding a "shoulder" which can be roughly detected
in the CDF13  data as well (which we name it MAK shoulders). In the
panels (f) and (g) of the figure \ref{fige6} and the panels (a), (b)
and (c) of the figure \ref{fige7}, we have compared our results to
the experimental data, regarding the $\Delta\Phi_{\gamma \gamma}$
dependency of the IPPP production rates. The
low-$\Delta\phi_{\gamma\gamma}$ tail can be clearly seen here.

It is interesting to note that in these relatively low CM energies,
the LO-MRW framework performs  much better with respect to other
schemes, specially in the D013 data. Additionally, one finds out
that in the higher transverse momenta, i.e. where the higher-order
pQCD effects become important, the KMR results behave similar to the
NLO-MRW rather than its LO counterpart and the KMR and MRW-NLO
results are below the experimental data. So their behaviors are the
same, while MRW-LO approximately cover the data. Generally speaking,
in the low-$\Delta\Phi_{\gamma \gamma}$ domain, the effect of the
fragmentation and the higher-order contributions are large (see the
panels (b), (d)  and  (e) of the figure \ref{fige6} and the panels
(a), (d)  and  (g) of the figure \ref{fige7}). Hence as a clear
pattern, the KMR and LO-MRW results are larger compared to the
NLO-MRW ones. Because of the different AOC implementations on these
prescriptions, the predictions get quite separated in their
respective regions. We should point out that the MRW sub-processes
in the above calculations behave roughly the same as those of KMR.
However, some discrepancy in case of NLO-MRW is observed. There is
not a sizable difference between the above schemes which use various
AOC and cut off in the differential cross sections, which is in
agreement to reference \cite{re}. As one should expect, this is not
the case on fragmentation domain.
\subsection{$E_{CM}=7\;TeV$}
\begin{figure}[H]
\includegraphics[scale=0.3]{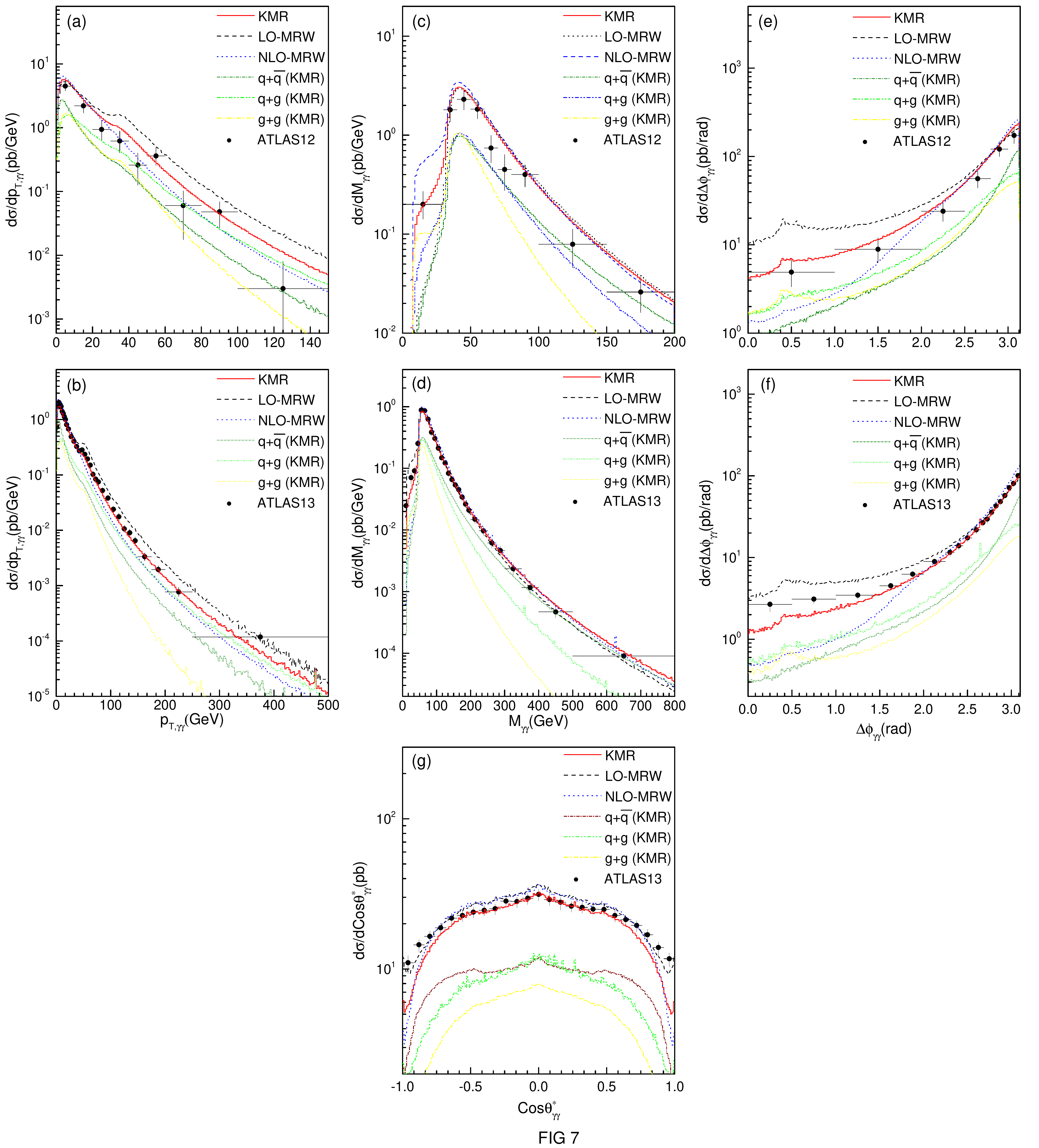}
\caption{The differential cross section of the production of $IPPP$
as functions of the transverse momentum ($p_{t,\gamma\gamma}$ in the
panels (a) and (b)), the photon invariant mass ($M_{\gamma\gamma}$
in panels (c) and (d)),  $\phi_{\gamma\gamma}$ (in the panels (e)
and (f) ) and $cos\theta_{t,\gamma\gamma}^{*}$ (in the panel (g)) at
$E_{CM}=7 \;TeV$. The experimental data are from the $ATLAS$
collaboration, \cite{ATLAS10,ATLAS13}. Note that the sub-processes
are only given for the KMR approach.}
 \label{fige8}
\end{figure}

\begin{figure}[H]
\includegraphics[scale=0.3]{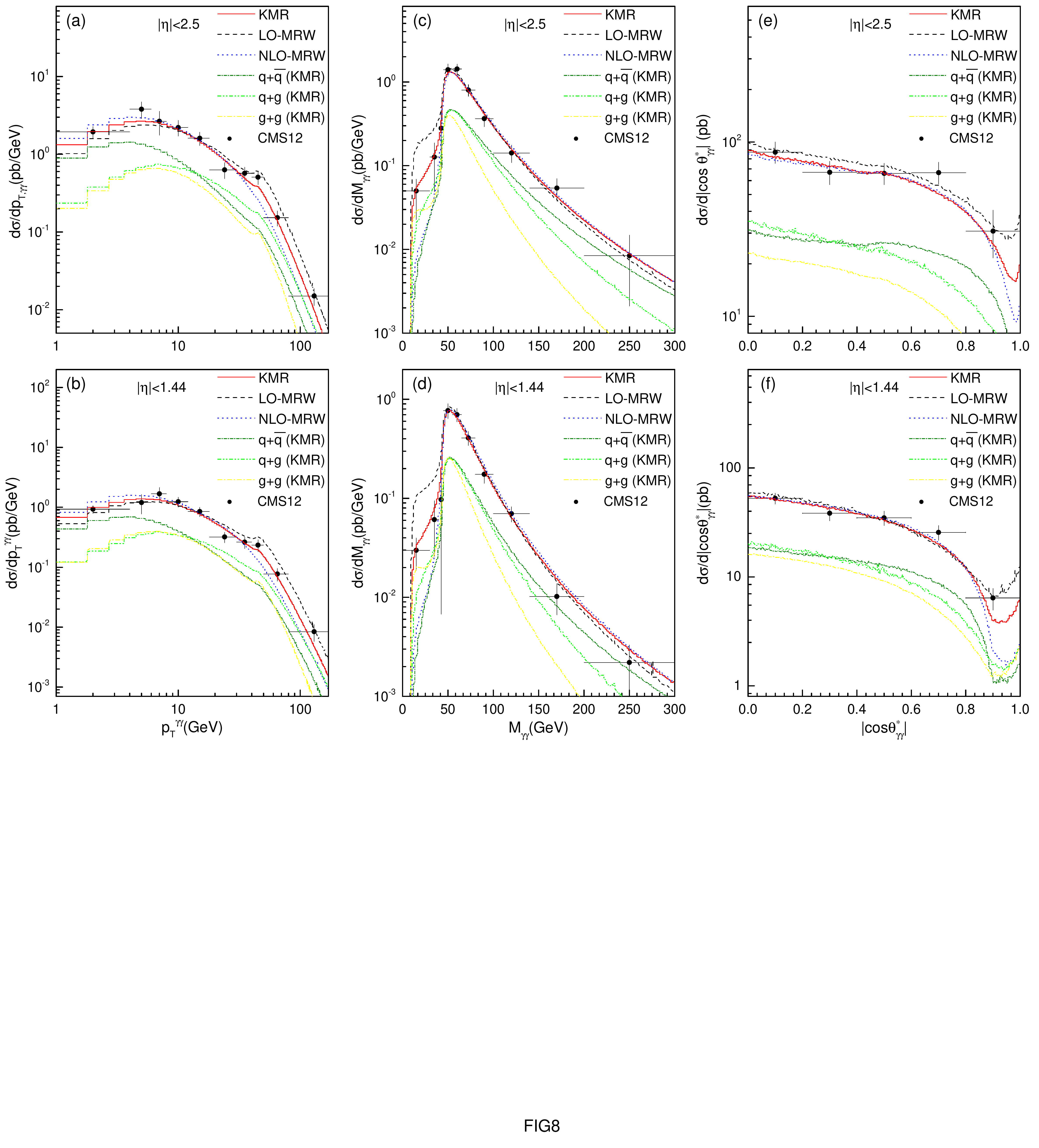}
\caption{The differential cross section of the production of $IPPP$
as functions of the transverse momentum ($p_{t,\gamma\gamma}$ in the
panels (a) and (b)), the photon invariant mass ($M_{\gamma\gamma}$
in the panels (c) and  (d)) and $cos\theta_{t,\gamma\gamma}^{*}$ (in
the panels (e) and  (f)) at $E_{CM}=7 \;TeV$. The experimental data
are from the CMS collaboration, \cite{CMS11}. Note that the
sub-processes are only given for the KMR approach.} \label{fige9}
\end{figure}

\begin{figure}[H]
\includegraphics[scale=0.3]{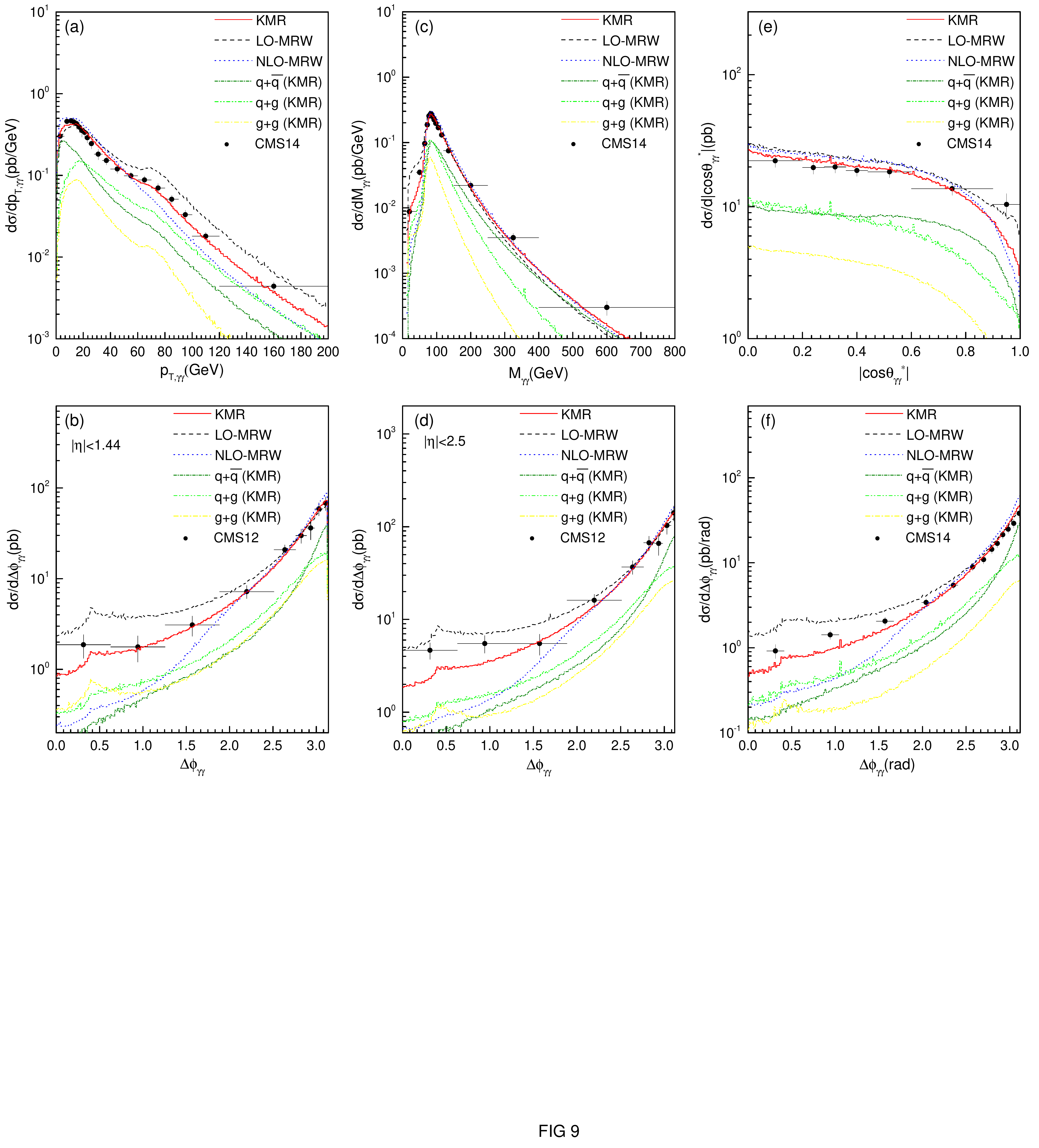}
\caption{The differential cross section of the production of $IPPP$
as functions of the transverse momentum ($p_{t,\gamma\gamma}$ in the
panel (a)),  the photon invariant mass ($M_{\gamma\gamma}$ in the
panel (c) ), $cos\theta_{t,\gamma\gamma}^{*}$ (in the panel (e)) and
$\Delta\phi_{\gamma\gamma}$ (in the panels (b), (d) and (f) ) at
$E_{CM}=7 \;TeV$. The experimental data are from the $CMS$
collaboration, \cite{CMS11,CMS14}. Note that the sub-processes are
only given for the KMR approach.} \label{fige10}
\end{figure}

We have performed another set of calculations, with the CM energy of
$7\;TeV$, in accordance with the specifications of the ATLAS and the
CMS reports, i.e. the references \cite{CMS11,CMS14,ATLAS10,ATLAS13}.
Therefore, in the figures \ref{fige8} through \ref{fige10}, the
reader is presented with comprehensive comparisons regarding the
dependency of the differential total cross-section of the IPPP
production, as the functions of $p_{t,\gamma\gamma}$,
$M_{\gamma\gamma}$, $\Delta\phi_{\gamma\gamma}$ and
$cos\theta^{*}_{\gamma\gamma}$. The general behavior of the results
are the same as in the $E_{CM}=1.96\;TeV$ case, with the exception
that  the contributions coming from the $q^*(\bar{q}^*) + g^* \to
\gamma + \gamma+ q(\bar{q})$ sub-process   is visibly greater than
that of the $q^* + \bar{q}^* \to \gamma + \gamma$ sub-process, with
some exceptions for the LO-MRW case (we do not present their data in
order not to crude these figures). This happens, because the shares
of the gluons and the sea-quarks become important, with the increase
of the CM energy.

As a result of increasing the CM energy, the Guillet shoulder
phenomena can be seen more clearly in our $7\;TeV$ calculations.
Additionally, as in the $1.96\;TeV$ case, in the regions where the
fragmentation effects become non-negligible, the low-tails of mass
and the low-$\Delta\Phi_{\gamma \gamma}$ tails are visible and
generally followed by the separation of the KMR, LO-MRW and NLO-MRW
results. Generally speaking, in these areas the LO-MRW and the
NLO-MRW results are the lower and the upper bounds
 relative to the KMR diagrams, respectively.

One should note that, the asymmetric constraint is applied on the
transverse energies of the IPPP production in the CMS14  data. So,
we
 perform our calculations for $q^*(\bar{q}^*)+g$ and $g^*+q^*(\bar{q}^*)$
configurations of the $q^*(\bar{q}^*) + g^* \to \gamma + \gamma+
q(\bar{q})$ sub-process, separately. As a result of this asymmetric
constraint, the production of the back-to-back photons are being
suppressed in the transverse plane \cite{CMS14}. Therefore, the
higher order contributions, e.g. the quark-gluon scattering, become
more important. In the CMS14   measurement, despite our
expectations, the quark annihilation has a significant contribution
in the LO-MRW, while the KMR and the NLO-MRW results have the
"expected" behavior. Nevertheless, only the KMR prescription is
somehow successful in describing the experimental data in "these
kinematic regions".

Unlike the CMS14  measurement, the ATLAS12
 experiment utilizes symmetric constraints \cite{ATLAS13}. Therefore as
 one
expects, the lower order pQCD contributions should be enhanced these
data.  So the NLO-MRW should perform  a better behavior for the
prediction of
 the experimental outcome, see for example the figure
\ref{fige8}. Therefore, we may  conclude that by including suitable
higher-order contributions, in accordance with the experiment
conditions, i.e. the imposed kinematics constraints,  the
predictions of the NLO-MRW framework becomes better and may be more
consistent with respect to two other $k_t$-factorization approaches.

\begin{figure}[H]
\includegraphics[scale=0.3]{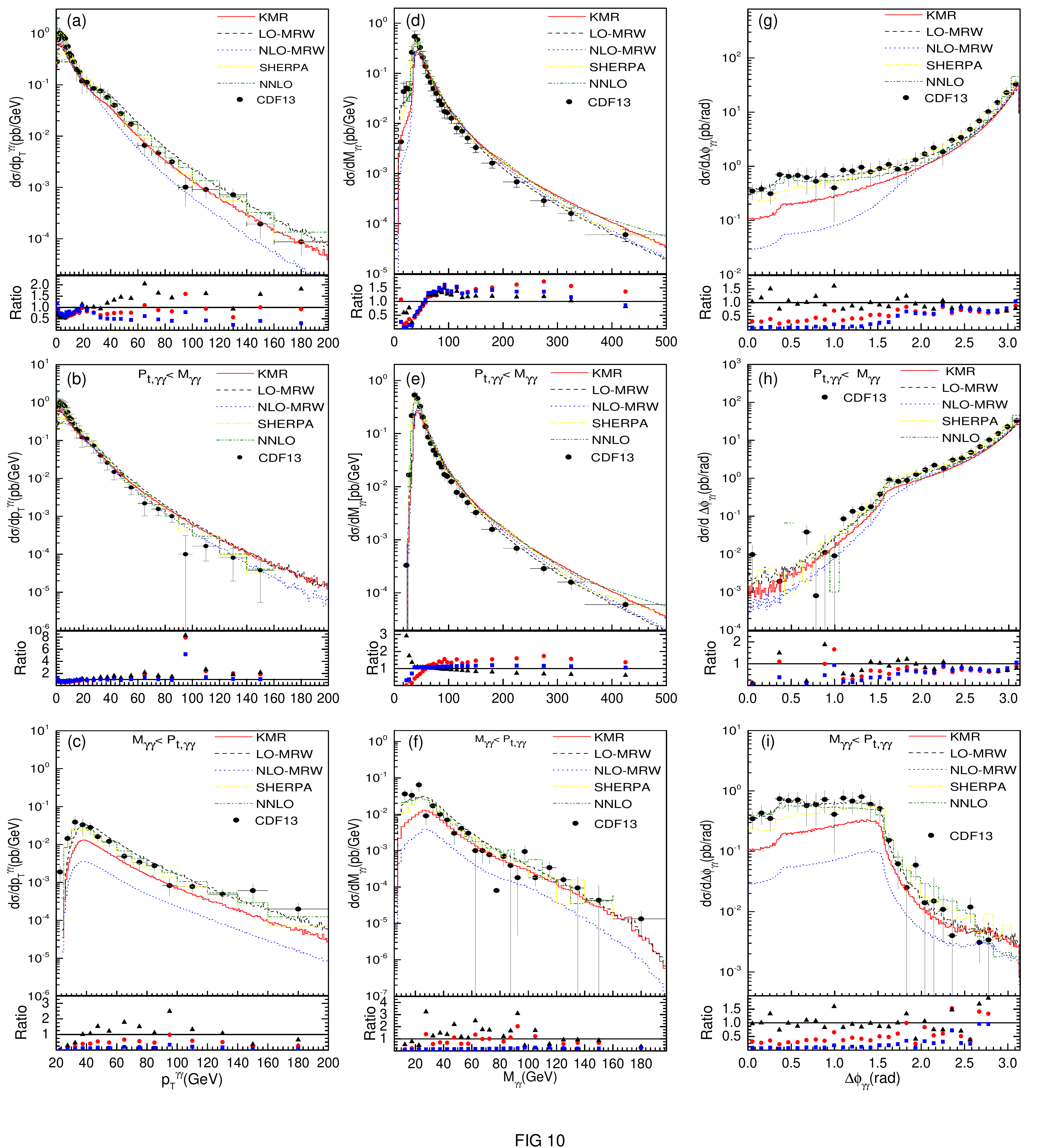}
\caption{The comparison of    $KMR$ and $MRW$ $IPPP$ differential
cross section (as function of $p_{t,\gamma\gamma}$,
$M_{\gamma\gamma}$ and $\Delta\phi_{\gamma\gamma}$ at $E_{CM}=1960
\;GeV$ )  with the $SHERPA$ and $NNLO$ $pQCD$ ($2\gamma NNLO$)
calculations base on the $CDF13$ experimental data \cite{CDF13}. The
lower panels present the ratio of our computation to  that of the
corresponding experimental data (the red circles, the black
triangles and the blue squares, show the ratio of KMR, LO-MRW and
NLO-MRW, respectively).}
 \label{fige11}
\end{figure}

\begin{figure}[H]
\includegraphics[scale=0.3]{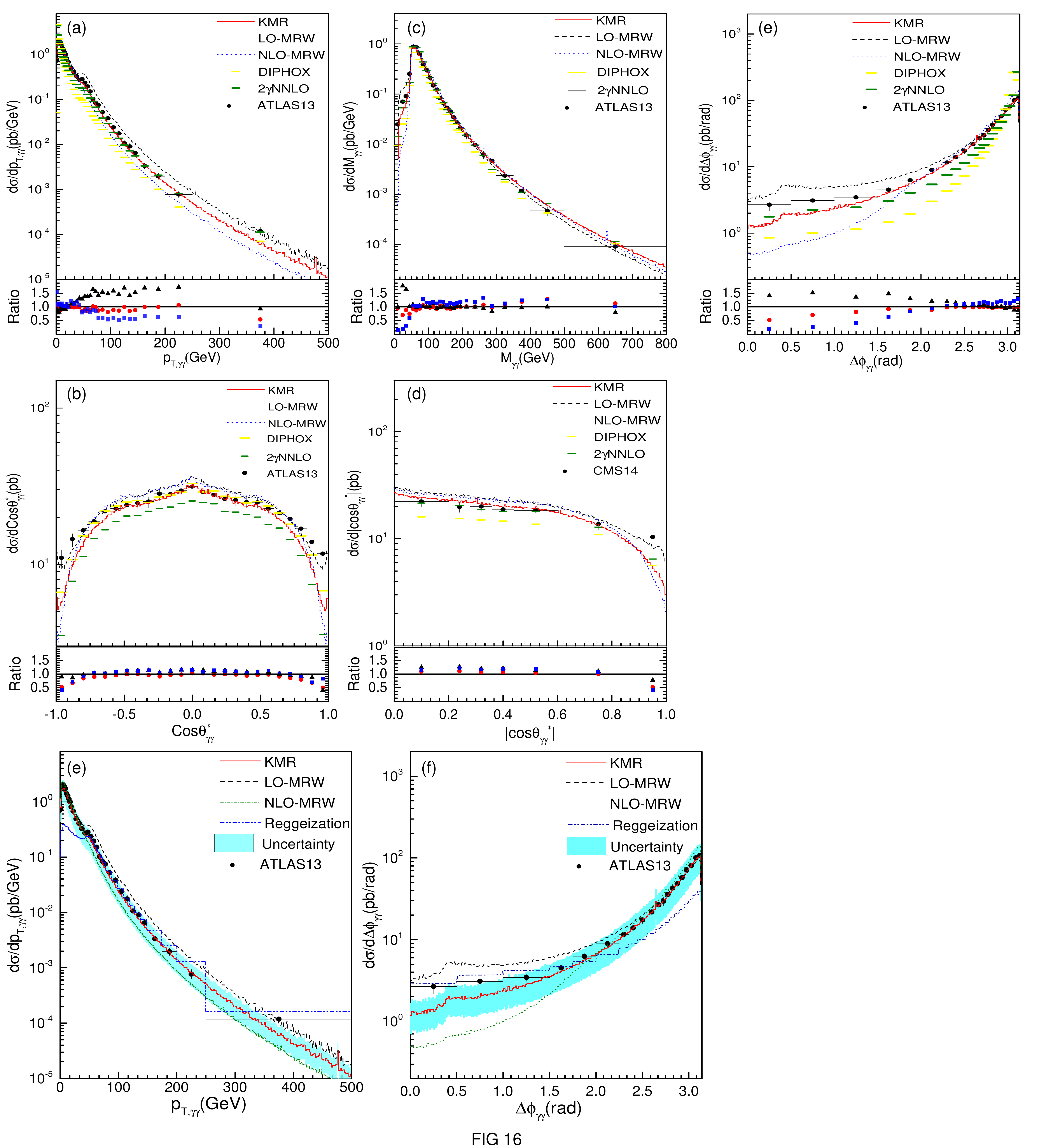}
\caption{As the figure 15 but for the $CMS14$ and $ATLAS13$
($E_{CM}=7 \;TeV$) experiments and the $2\gamma NNLO$  and $DIPHOX$
results . The panels (f) and (g) are similar to those of (a) and (e)
but with the reggeization method of reference \cite{saleev1} and the
uncertainty bands for our calculation. }
 \label{fige17}
\end{figure}

Finally, we would like   to present a comparison   between our
results and the Monte Carlo event generator as well as the  pQCD
which were introduced in the  introduction
\cite{RESBOS,DIPHOX,NNLO,sherpa}. Furthermore, we make  a careful
scrutiny of our calculation  by   dividing     our  results in
different frameworks  (i.e. KMR, LO-MRW and NLO-MRW)  to  that of
the corresponding  experimental data. This can highlight the
difference of our works over the experiments. The outcome of above
comparisons are demonstrated in  the figure \ref{fige11} through
figure \ref{fige17}.  At the lower panels of these figures the red
circles show the KMR ratio and the black triangles and  the blue
squares are presenting  the LO-MRW and NLO-MRW  ratios as explained
above, respectively.

\begin{figure}[H]
\includegraphics[scale=0.3]{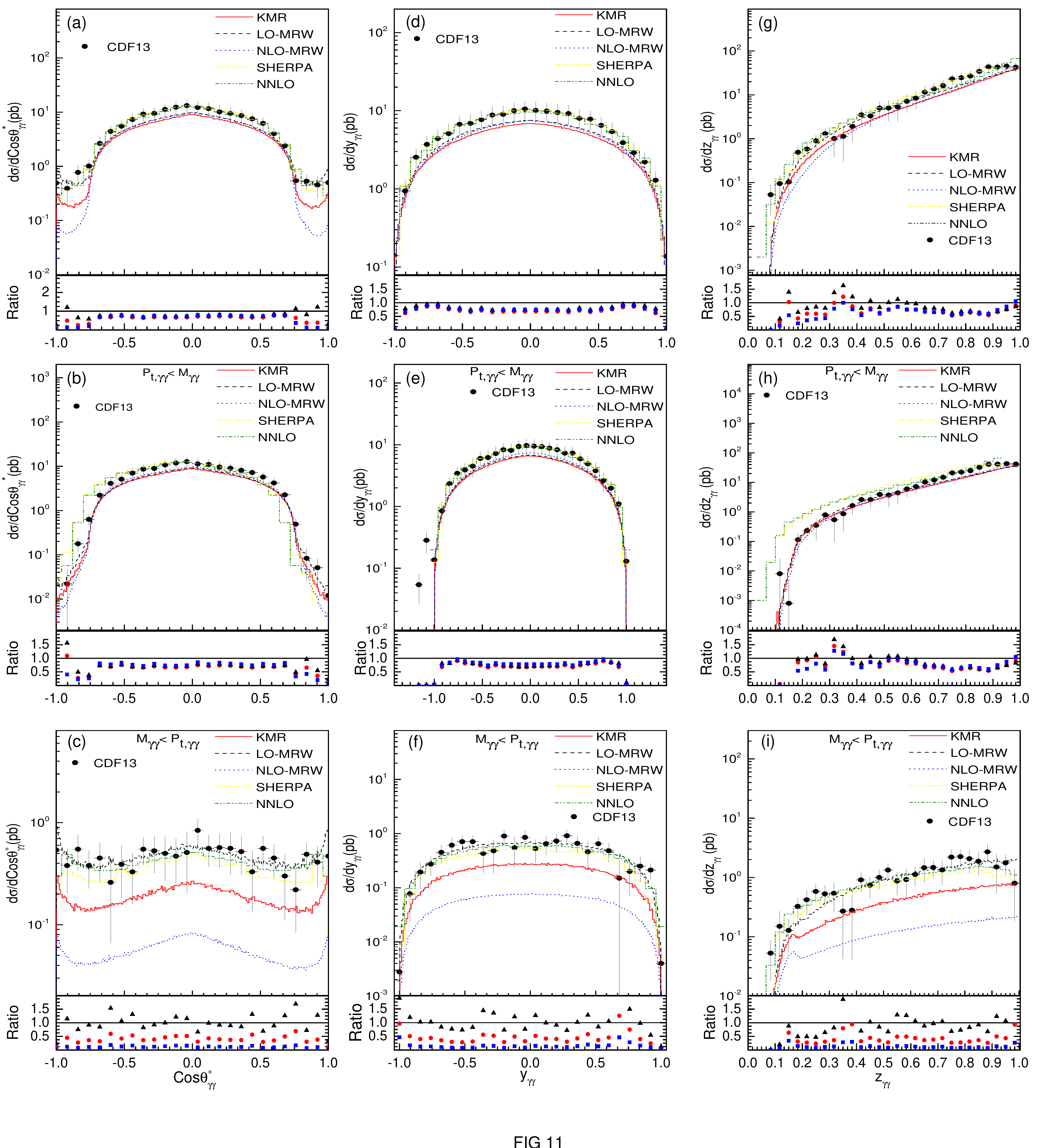}
\caption{The same as  the figure 11, but for the
$cos\theta_{t,\gamma\gamma}^{*}$, $y_{\gamma\gamma}$ and
$Z_{\gamma\gamma}$ variables.}
 \label{fige12}
\end{figure}

\begin{figure}[H]
\includegraphics[scale=0.3]{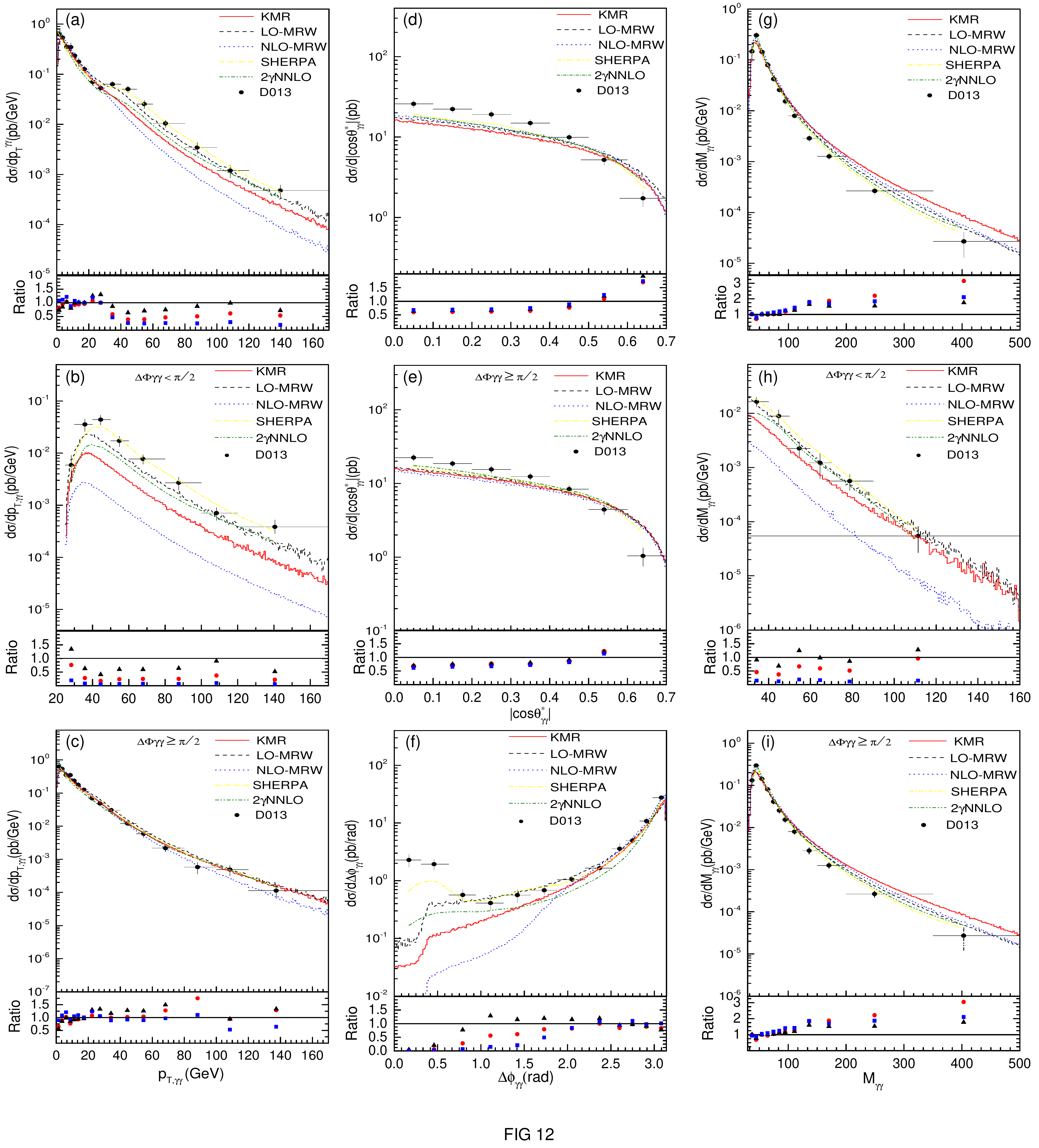}
\caption{The same as the figure 11 but for the $D013$ experiment and
extra $cos\theta_{t,\gamma\gamma}^{*}$ variable.}
 \label{fige13}
\end{figure}

In the   figures \ref{fige11}, \ref{fige12} and \ref{fige13}, our
KMR ,LO-MRW and NLO-MRW results are compared with the SHERPA and the
NNLO (or $2\gamma$NNLO)  pQCD \cite{CDF13,D013}, as well as CDF13
and D013 data. It is observed that our KMR approach predicts the
differential cross section data as a function of corresponding
variable very well, but in the region where the fragmentation is not
important. While in the regions where the fragmentation and higher
order pQCD become dominant  (for example, low $\Delta\phi$ region in
the panel (g) of the figure \ref{fige11}), the SHERPA and NNLO
methods produce better results. At these regions without considering
higher order contributions and fragmentation, only the MRW can
predict experimental data correctly, especially this well behavior
can be observed in the panels (c), (f) and (i) of the figures
\ref{fige11} and \ref{fige12} and the panels (b) and (h) of the
figure \ref{fige13}. However, the SHERPA and $2\gamma$NNLO
calculations are well behaved in whole regions and all of the
panels.

\begin{figure}[H]
\includegraphics[scale=0.3]{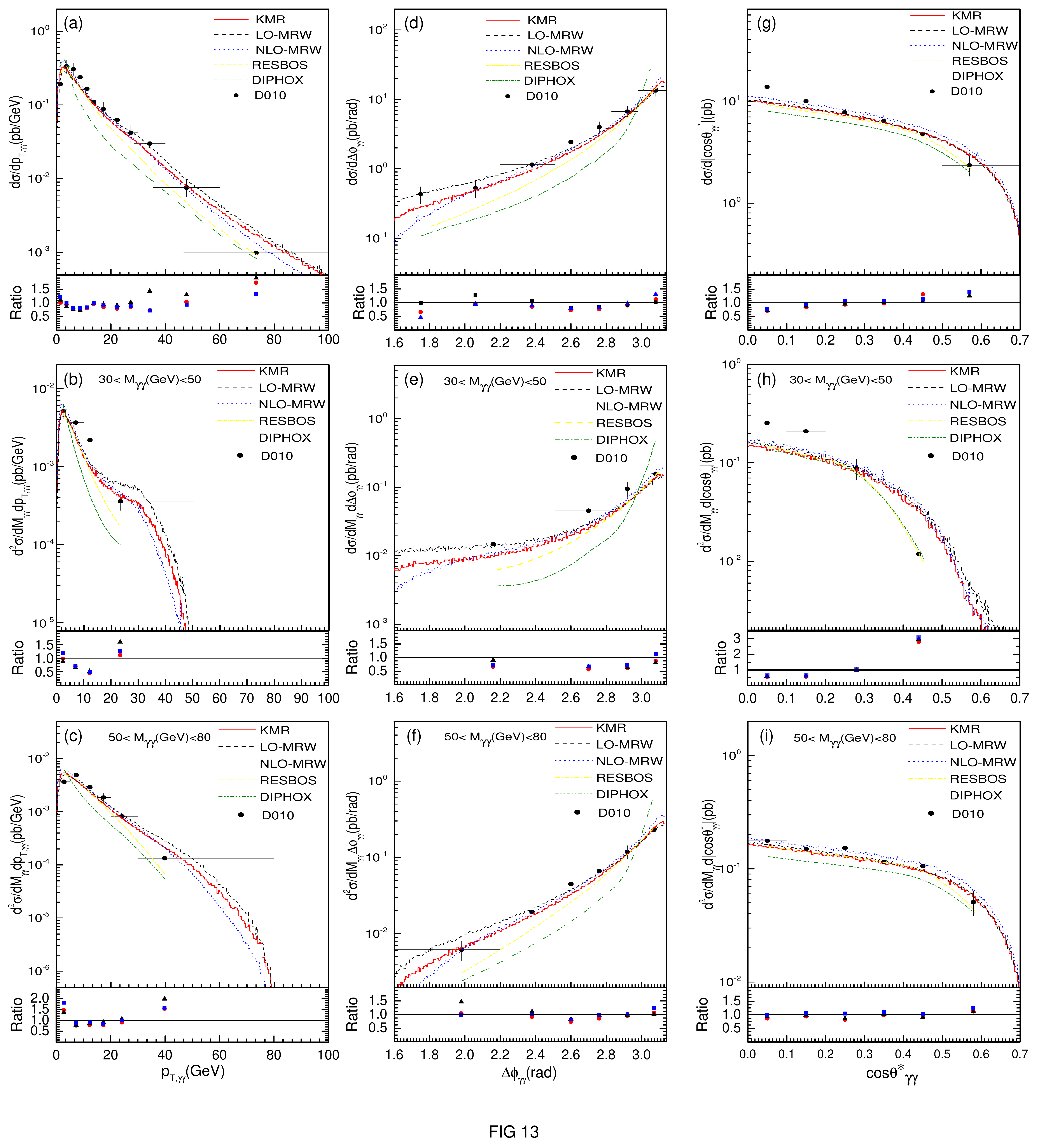}
\caption{The same as the figure 11 but with the RESBOS and DIPHOX
results and $p_{t,\gamma\gamma}$, $\Delta\phi_{\gamma\gamma}$ and
$cos\theta_{t,\gamma\gamma}^{*}$, variables.}
 \label{fige14}
\end{figure}

\begin{figure}[H]
\includegraphics[scale=0.3]{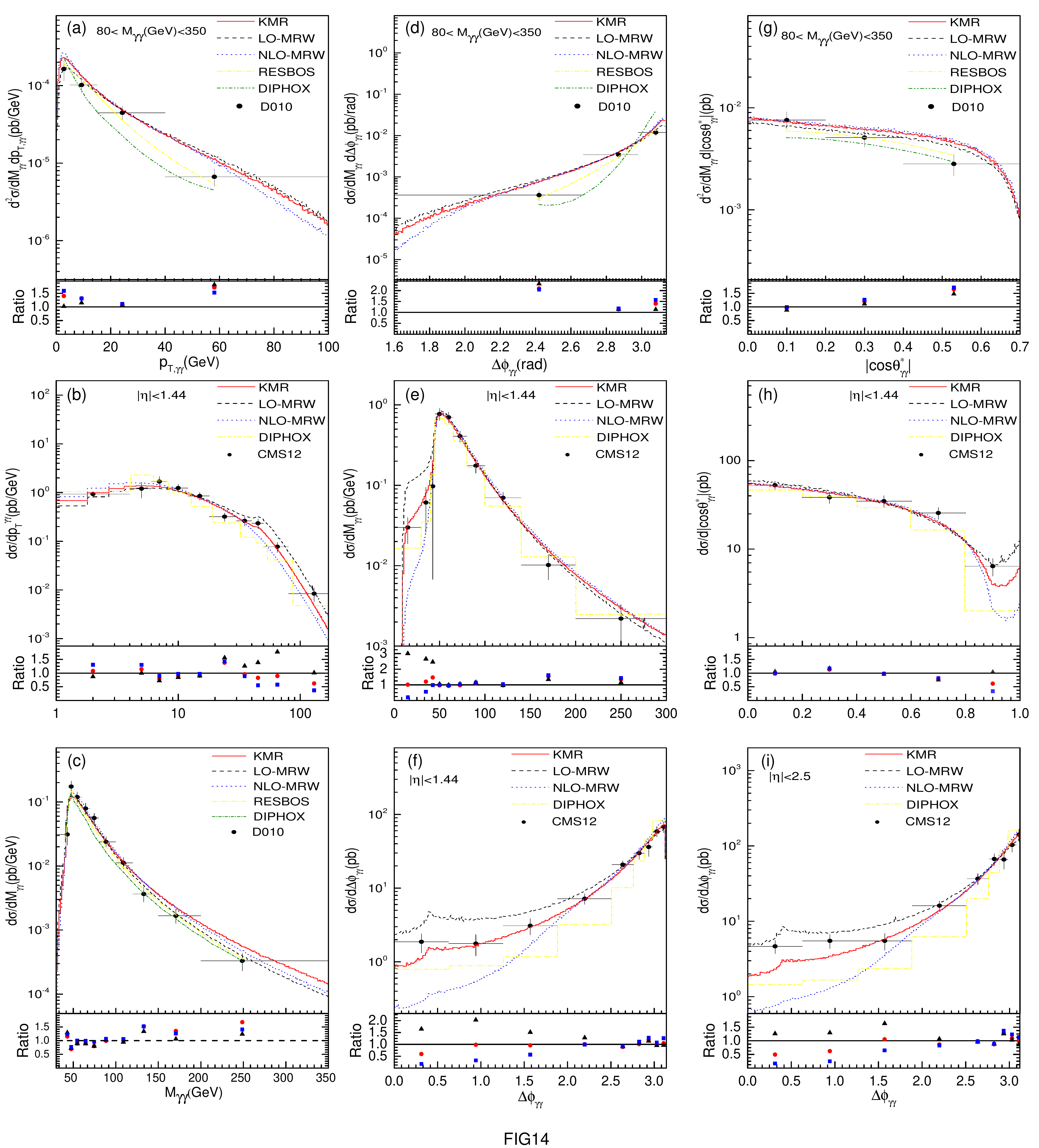}
\caption{As the figure 14 but for the D010 ($E_{CM}=1960 \;GeV$)and
CMS12 ($E_{CM}=7 \;TeV$) experiments and extra $M_{\gamma\gamma}$. }
\label{fige15}
\end{figure}

The figure \ref{fige14}  compares the D010  data \cite{D010} with
our results, as well as  the RESBOS and DIPHOX  calculations
\cite{D010}. One can clearly observe that the  KMR
$k_t$-factorization approach predicts the acceptable result with
respect to other theoretical methods that presented in this figure,
especially for all of  the double differential cross section
channels (i.e. the panels (b) to (i)). On the other hand, in the
panels (a), (d) and (g) of the figure \ref{fige15} and in the high
value of $M_{\gamma\gamma}$, the Monte Carlo calculation is more
successful. The remaining panels of this figure which is related  to
the CMS
 collaborations \cite{CMS11,CMS14}, show that our results predict the data   with
higher accuracy  with respect to those of DIPHOX calculation
\cite{CMS11,CMS14,DIPHOX}.

\begin{figure}[H]
\includegraphics[scale=0.3]{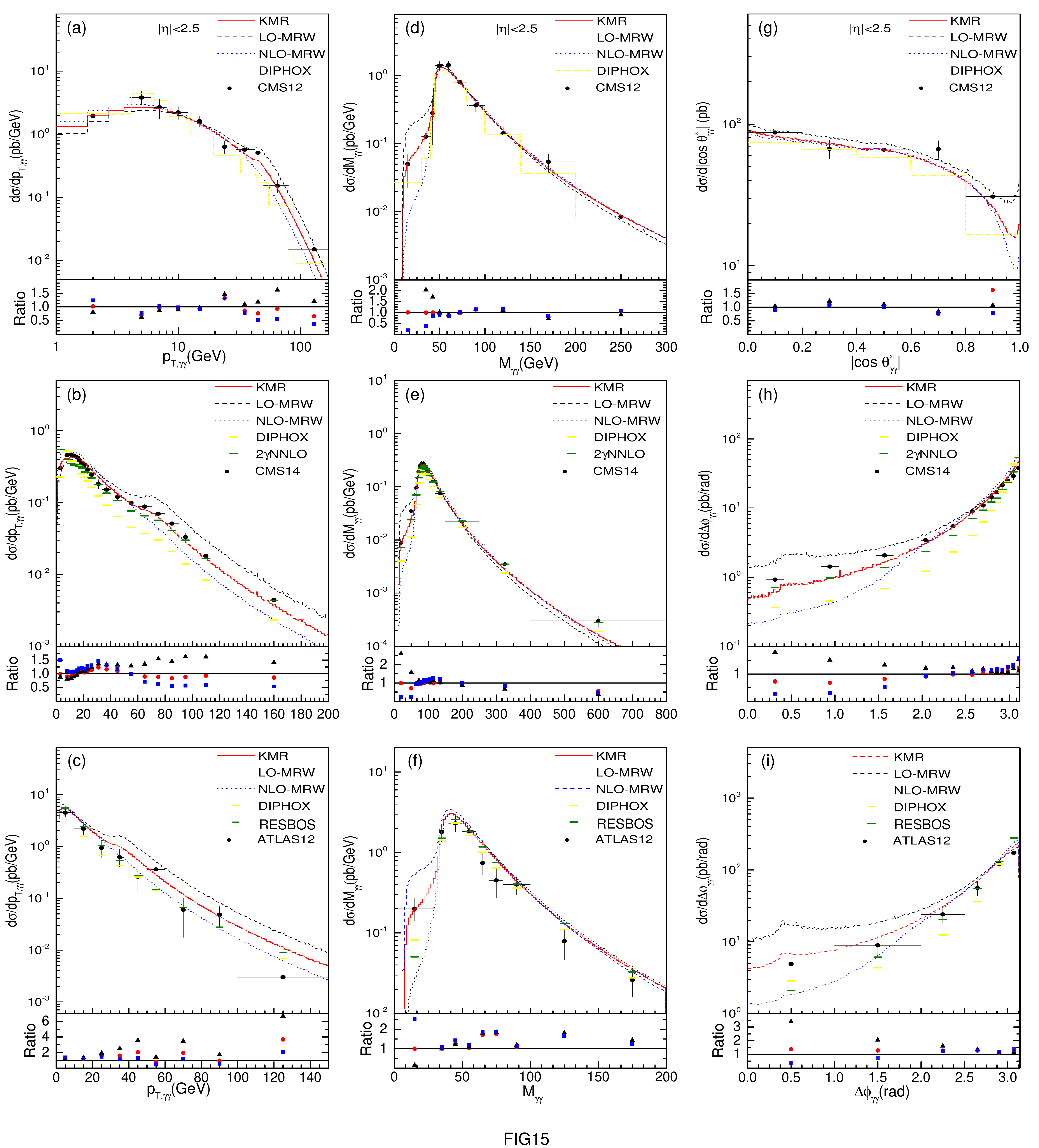}
\caption{As the figure 15 but for the CMS12, CMS14 and ATLAS12
($E_{CM}=7 \;TeV$) and with the $2\gamma$NNLO, $RESBOS$ and $DIPHOX$
results.}
 \label{fige16}
\end{figure}

In the panels (a), (d) and (g) of  the figure \ref{fige16} similar
to the figure \ref{fige15}, the KMR approach behaves as before,  but
the results of DIPHOX calculation are closer to the data, since the
rapidity was increased. In rest of the panels of the figures
\ref{fige16} and  the panels (a) to (e) of \ref{fige17}, our results
are examined against DIPHOX and $2\gamma$NNLO, and their behavior
are much similar. However, our KMR or MRW, as it was discussed
before, are closer to the $2\gamma$NNLO calculation.

In order to check our results against those of reggization methods,
\cite{saleev,saleev1}, and also to give the uncertainty of present
calculation (by multiplying the factorization scale by half and two)
, the panels (a) and (e) are repeated in the panels (f) and (g)  of
the figure \ref{fige16}.  It is seen that the data are very close to
 those of reference \cite{saleev1}, except  the small
$p_{t,\gamma\gamma}$ and the large $\Delta\Phi_{\gamma\gamma}$
regions, where their results are off the data. However our
uncertainty bounds are reasonably cover the data as well as the
reggization and $2\gamma$NNLO calculations (see panels (a), (e), (f)
and (g)). It is interesting that the $2\gamma$NNLO method, in which
the fragmentation contribution has been also taken into account, is
off the experimental data while the reggization method   cover them.
As we pointed out in the end of section IV, beside the separation
and isolation cone conditions for possible double counting, we also
modified our UPDF according to e.g. the reference \cite{LIP2011} and
find less than 15 per cent effect, which still keeps our result
inside the uncertainty bounds. But, as we pointed out in the
introduction,  the UPDF should satisfy the condition given in the
equation (1), so any changes in the UPDF certainly affect the
original PDF definitions or it may be in contrast with the original
definition of UPDF \cite{KIMBER}. On the other hand, there is no
grantee that the $k_t$ factorization method produces results  better
than those of pQCD as it is stated by Martin et al \cite{WattWZ}.

These comparisons   show that one of the places in which the effect
of $k_t$ factorization framework obviously becomes important, with
respect to its counterpart, would be the regions of large
$\Delta\phi_{\gamma\gamma}$ and small $p_{t,\gamma\gamma}$. In these
two regions, the predictions of collinear matrix element method are
overestimated the data, as it could be seen by DIPHOX  and
$2\gamma$NNLO calculations. However, the prediction of RESBOS, due
to the NLL (next leading logarithmic) resumption of soft initial
state gluon radiation, is better than those of DIPHOX  and
$2\gamma$NNLO. But in   our methods, the natural  gluon resumption
automatically is done in all orders \cite{SPL}, because of the
Sudakov form factor, so this problem would not exist.
\section{conclusions}
Throughout this work, we   calculated the rate of the production of
the isolated prompt-photon pairs, in the $k_t$-factorization
framework, using the UPDF of the KMR and the MRW prescriptions and
compared our results to the existing experimental data from the D0,
CDF, CMS and ATLAS collaborations. According to our discussions and
observations  in the present work, the LO-MRW approach is the best
suitable scheme for the prediction of the IPPP production rates in
the lower CM energies, since this approach can predict the
experimental data within the regions where the fragmentation effects
become important, without any additional manipulations in our
calculations. In contrary, the LO-MRW formalism is not perfect for
the higher CM energies in these kinematics. While the KMR approach
is able to accurately predict the experimental data in the 7 TeV
center of mass energy. The main difference between these approaches
arises due to the implementation of different visualizations of the
AOC, which can be seen, specially in the regions where the
fragmentation and the higher-order contributions become important,
i.e. when the quark-radiation terms are enhanced. In these areas, we
expect that the three approaches behave well-separated. On the other
hand it was shown that the application of different  AOC and  cut
off, using the KMR \cite{re} and MRW prescriptions do not show
serious discrepancies and a qualitative agreements between different
schemes can be achieved.

We   realized that the  Guillet shoulder phenomena is more sensitive
to the low-$M_{\gamma\gamma}$ variations, compared to the
low-$\Delta\phi_{\gamma\gamma}$ and the $cos\theta^*_{\gamma\gamma}$
regions. Although, our predictions via our simplistic calculations
describe the experimental data well,  one can improve the precision
of these results by including higher-order contributions and taking
into account the fragmentation effects. We hope that in our future
works, we can investigate these phenomena. A comparison was also
made with  the different theoretical methods such as DIPHOX ,
$2\gamma$NNLO, RESBOS and SHERPA and an overall agreement was found.

It was shown that the possible double-counting can be removed by
considering the phase space cuts as well as modification in the
UPDF. However by imposing  the uncertainty of the factorization
scale in the resulted differential cross section, this issue may not
be important.

In this work we used the small x approximation, however as we stated
in the introduction, we can use the effective action approaches for
the off shell partons. We hope to investigate this approximation in
our future works \cite{KUTAK1,KUTAK2,KUTAK3,KUTAK4,LIP2016} as well
as the gauge invariance and possible double counting.
\begin{acknowledgements}
MM would like to acknowledge the Research Council of University of
Tehran and the Institute for Research and Planning in Higher
Education for the grants provided for him. RAN and MRM sincerely
thank  M. Kimber and A. Lipatov for their valuable discussions and
comments.
\end{acknowledgements}



\appendix
\section{The constraints of various experiments}
We provide our results, by considering all constraints
that are imposed in each experiment, among which  two of them are
considered generally, i.e. the isolated-cone and the separation-cone
constraints. The isolated-cone is responsible for distinguishing the
"non-prompt decay photons" from the prompt-photons. This constraint
requires the transverse energy $E_t^{iso}$ (in a cone with the
angular radius   $
R=\sqrt{(\eta-\eta^\gamma)\;+\;(\phi-\phi_\gamma)} \;\;<\;0.4 $) to
be less than a few GeV according to each experiment. To avoid, the
overlap between the two photons, the separation-cone constraint is
imposed as,
$$
\Delta R\;>\;0.4.
$$
 Other constraints such as the $p_t$-threshold of prompt-photons, the pseudo-rapidity
 regions, etc, are imposed  according to the settings of the individual
 experiments. Obviously, the different settings probe the various regions of pQCD. In
 what follows, we briefly present the reader with the specifications of the
 measurements that we intend to analyze throughout this work in each experiments.
\subsection{The D0 collaboration}
The D0 experiment was performed at the center of mass energy of
$1.960$ TeV. The two sets of D0 data, related to the IPPP production
were investigated in the references \cite{D010,D013}, i.e. D010 and
D013, respectively. The constraints in the D010 experiment
\cite{D010} are $p_t$ $>$ 20, 21 \;GeV (the transverse momentum of
outgoing photons),
 $|\eta|<0.9$ (the pseudo-rapidity) and $M_{\gamma\gamma}\;>\;p_{t,\gamma\gamma}$  which are applied
 to the IPPP production \cite{D010}, suppressing the fragmentation effects and some higher order
  contributions. In the D013 report \cite{D013}, the constraints are  $p_t$ $>$ 17, 18 $GeV$ and
  $|\eta|<0.9$. Also three regions are probed in the D013  report, i.e the region I
  with the $\Delta\phi_{\gamma\gamma} \geq {\pi\over 2}$ constraint which is suitable for
   the study of non-higher-order pQCD. By applying the $\Delta\phi_{\gamma\gamma} < {\pi\over 2}$
   constraint in the region II, the fragmentation effects become important and the last region
   is without any extra constraint on $\Delta\phi_{\gamma\gamma}$ \cite{D010,D013}.
\subsection{ The CDF  collaboration}
Similar to the D0 collaboration, the CDF experiment provides the two
sets of data, that are related to  the IPPP production
\cite{CDF11,CDF13} at the center of mass energy of $1.960$ TeV. The
constraints in the CDF13 \cite{CDF13} are the same as the CDF11
\cite{CDF11} reports. However, the luminosity is improved in the new
sets of data (CDF13) \cite{CDF13} . These constraints are $p_t$ $>$
15, 17 GeV and $|\eta|<1$. The $CDF$ collaboration explored three
regions in their works: the region I, via applying the
$p_{t,\gamma\gamma} > M_{\gamma\gamma}$ constraint, suitable for the
study of higher-order pQCD. The region II, by applying the
$p_{t,\gamma\gamma} < M_{\gamma\gamma}$ constraint, which undermines
the fragmentation effects and emphasizes on the quark-antiquark
annihilation. The last region is without any extra constrains
\cite{CDF11,CDF13}.
\subsection{The CMS collaboration}
The CMS collaboration is presented  at the  7 TeV CM energy
\cite{CMS11,CMS14}. The constraints in the CMS12 experiment
\cite{CMS11} are $p_t$ $>$ 20, 23 GeV and $|\eta|<2.5$, excluding
the $1.44<|\eta|<1.57$ region. Also, the region $|\eta|<1.44$ is
separately canalized. In the reference  \cite{CMS14} ($CMS14$)  the
asymmetric transverse momentum  ($p_t$ $>$ 40, 25 GeV ) for  the
IPPP production is selected in the regions of $1.57<|\eta|<2.5$ and
$|\eta|<1.44$. As a result, the higher order pQCD contributions
become dominant in these experiments \cite{CMS11,CMS14}.
\subsection{The ATLAS collaboration}
Another experimental data at the 7 TeV CM energy is provided by the
ATLAS collaboration \cite{ATLAS10,ATLAS13} (ATLAS12 and ATLAS13). In
the reference \cite{ATLAS10}, the data is sorted according to $p_t$
$>$16 (16) GeV and $|\eta|<2.5$, excluding the $1.37<|\eta|<1.52$
region. Similarly, ATLAS13 \cite{ATLAS13} has the same
pseudo-rapidity region, although the transverse momentum threshold
is changed to $P_t$ $>$25 (22) GeV.
\end{document}